\documentclass[fleqn,usenatbib]{mnras}

\usepackage{newtxtext,newtxmath}

\usepackage[T1]{fontenc}

\DeclareRobustCommand{\VAN}[3]{#2}
\let\VANthebibliography\thebibliography
\def\thebibliography{\DeclareRobustCommand{\VAN}[3]{##3}\VANthebibliography}

\usepackage{graphicx}	
\usepackage{amsmath}	
\usepackage{float}
\usepackage{natbib}
\usepackage{caption}
\usepackage{gensymb}
\usepackage{bm}
\usepackage[output-decimal-marker={,},exponent-product=\cdot]{siunitx}
\usepackage[utf8]{inputenc}
\usepackage{newtxtext,newtxmath}
\usepackage{hyperref}

\title[Buildup of galaxies and their spheroids]{The buildup of galaxies and their spheroids: the contributions of mergers, disc instabilities and star formation}

\author[F. Huško, C. G. Lacey \& C. M. Baugh]{
Filip Huško,$^{1}$\thanks{E-mail: filip.husko@durham.ac.uk}
Cedric G. Lacey,$^{1}$
Carlton M. Baugh$^{1}$
\\
$^{1}$ Institute for Computational Cosmology, Department of Physics, University of Durham, South Road, Durham, DH1 3LE, UK\\
}

\date{Accepted XXX. Received YYY; in original form ZZZ}

\pubyear{2022}

\begin{document}

\label{firstpage}
\pagerange{\pageref{firstpage}--\pageref{lastpage}}
\maketitle

\begin{abstract}
We use the {\tt GALFORM} semi-analytical model of galaxy formation and the Planck-Millennium simulation to investigate the origins of stellar mass in galaxies and their spheroids. We compare the importance of mergers and disc instabilities, as well as the starbursts that they trigger. We find that the fraction of galaxy stellar mass formed \textit{ex situ} (i.e. through mergers; $f_\mathrm{ex}$) increases sharply from $M_*=10^{11}$ M$_\odot$ upwards, reaching $80\%$ at $M_*=10^{11.3}$ M$_\odot$. The massive end of the $f_\mathrm{ex}$-$M_*$ relation does not evolve with redshift, in disagreement with other models. For low-mass galaxies we find larger \textit{\textit{ex situ}} contributions at $z=0$ than in other models ($7$-$12\%$), with a decrease towards higher redshifts. Major mergers contribute roughly half of the \textit{ex situ} mass, with minor mergers and smooth accretion of satellites both accounting for  $\approx25\%$, almost independent of stellar mass and redshift. Mergers dominate in building up high-mass ($M_\mathrm{*,sph}>10^{11}$ M$_\odot$) and low-mass ($M_\mathrm{*,sph}<10^{8.5}$ M$_\odot$) spheroids. Disc instabilities and their associated starbursts dominate for intermediate-mass spheroids ($10^{8.5}<M_\mathrm{*,sph}<10^{11}$ M$_\odot$) at $z=0$. The mass regime where pseudobulges dominate is in agreement with observed pseudobulge fractions, but the peak value in the pseudobulge fraction predicted by {\tt GALFORM} is likely too high. Starbursts induced by disc instabilities are the dominant channel for spheroid growth at all redshifts, while merger-induced starbursts are relatively negligible, except at very high redshifts ($z>5$).
\end{abstract}





\section{Introduction}

Galaxy mergers play an important role in the formation and evolution of galaxies. Most significantly, they provide a channel of stellar mass growth of galaxies alongside star formation, as well as transform galaxy discs into spheroids (\citealt{ToomreToomre1972},~\citealt{Schweizer},~\citealt{Barnes1992},~\citealt{Mihos}). They are also known to trigger starbursts (\citealt{Schweizer87},~\citealt{Barnes1991},~\citealt{Mihos1994},~\citealt{Hernquist1995}), affect the gas fraction of galaxies (\citealt{Larson2016},~\citealt{Kaneko2017},~\citealt{Ellison2018},~\citealt{Pan2018},~\citealt{Violino2018}), change the distribution and kinematics of stars (e.g.~\citealt{Naab2009},~\Citealt{vanDokkum2010},~\citealt{Newman2012},~\citealt{Ferreras2014}), and impact the growth of supermassive black holes (e.g.~\citealt{Treister2012},~\citealt{Rosario2015},~\citealt{Ellison2019}). 

Results from models of galaxy formation suggest that galaxy mergers are the main mode of growth for very massive galaxies ($M_*>10^{11}$ M$_\odot$; \citealt{Guo2008},~\citealt{Parry2009},~\citealt{Lacey2016}), while being fairly negligible for lower-mass galaxies. These findings are in line with observations of massive galaxies in clusters (e.g.~\citealt{McIntosh2008},~\citealt{Ferreras2014}). In recent years, theoretical studies have quantified the impact of mergers on galaxy growth through the \textit{ex situ} fraction (\citealt{RodrGomez2016},~\citealt{Dubois2016},~\citealt{LeeYi2017},~\citealt{Qu2017},~\citealt{Henriques2019},~\citealt{Davison},~\citealt{Moster2020}). This quantity represents the fraction of stellar mass that galaxies have accreted (as opposed to the stellar mass formed \textit{in situ} through star formation). Current galaxy formation models predict a qualitatively similar dependence of the \textit{ex situ} fraction on stellar mass in the local Universe. In particular, they find that \textit{in situ} star formation dominates in low-mass galaxies, while mergers dominate in high-mass ones. The mass which marks the transition between these regimes is found to be $M_*\approx10^{11}$ M$_\odot$ at $z=0$, although the exact details of this transition differ from model to model (e.g. \citealt{RodrGomez2016} and \citealt{Davison}). Furthermore, \textit{ex situ} fractions of low-mass and very high-mass galaxies are not the same in every model (e.g. \citealt{Dubois2016} and \citealt{Henriques2019}). In addition, some differences arise with increasing redshift (e.g. \citealt{Qu2017} and \citealt{Gupta2020}), but this is not surprising since most models are tuned to $z=0$ observations.

Galaxy mergers are very important in the creation and buildup of galaxy spheroids. Their role in this is apparent for very massive galaxies (e.g.~\citealt{McIntosh2008},~\citealt{Bundy2009}), since these galaxies are spheroid-dominated. For other galaxies, mergers compete with disc instabilities in the transformation of discs into spheroids. These gravitational instabilities occur due to the self-gravity of the discs (\citealt{Toomre1964},~\citealt{Ostriker1973}). 

In morphological terms, the population of spheroids can roughly be split onto classical spheroids and pseudobulges (\citealt{Kormendy2004}, \citealt{Obreja2013}, \citealt{Kormendy2013}, \citealt{MendezAbreu2014}). The former are defined by features which make them likely to be the result of mergers, while the latter are thought to have been created largely through disc instabilities. This interpretation is supported by simulations which predict bar formation as a result of disc instabilities (\citealt{Combes1981},~\citealt{Efstathiou1982},~\citealt{Raha1991},~\citealt{Christodoulou1995},~\citealt{Norman1996}). These bars are themselves unstable in the vertical direction and eventually thicken to form pseudobulges (\citealt{Combes1990},~\citealt{Pfenniger1990},~\citealt{Debattista2006},~\citealt{Gerhard2015},~\citealt{Erwin2016}). Other simulations and observations have found that discs can become stable by forming `clumps', which then migrate to the centres of galaxies due to dynamical friction, and form spheroids there (\citealt{Immeli2004},~\citealt{Elmegreen2008},~\citealt{Guo2015}). However, it is not clear if the spheroids formed in this manner are more like classical spheroids (e.g. \citealt{Elmegreen2008} or more like pseudobulges (e.g. \citealt{Inoue2012}) in terms of their characteristics. Semi-analytical models usually do not try to distinguish between different modes of pseudobulge creation.

It is possible for galaxies to have both a classical spheroid and a pseudobulge (\citealt{MendezAbreu2014},~\citealt{Erwin2015}). Pseudobulges have a more disc-like morphology (\citealt{Fisher2008},~\citealt{Gadotti2009},~\citealt{Irodotou2019},~\citealt{Luo2020}) and are rapidly rotating (e.g. \citealt{KormendyFisher2008}). They are largely hosted by intermediate-mass galaxies with significant disc components (\citealt{Fisher2011},~\citealt{Shankar2013},~\citealt{Vaghamere2013},~\citealt{Erwin2015}) and they have a much larger star formation rate (hereafter SFR; \citealt{Jogee2005},~\citealt{Fisher2009},~\citealt{Obreja2013}). The high SFR can be linked to ongoing starbursts triggered by disc instabilities (\citealt{Lehnert1996},~\citealt{Romeo2016},~\citealt{Tadaki2018}). Starbursts can also be triggered by mergers (e.g. \citealt{Schweizer87}, \citealt{Mihos1994},~\citealt{Cibinel2019},~\citealt{Patton2020}), but it is likely that the typical SFR in starbursts triggered by the two processes are different (as we show in Section~\ref{sec:StarFormationBursts}). Starbursts caused by one of both of these processes are thought to be especially important at high redshifts (\citealt{Guedes2013},~\citealt{Bournaud2016}).

Observations of galaxy spheroids find that pseudobulges are most often found in intermediate-mass galaxies, $M_*\approx10^{10}$ M$_\odot$ (\citealt{Fisher2011},~\citealt{Vaghamere2013},~\citealt{Erwin2015}). The fraction of galaxies hosting a pseudobulge is found to peak around this mass, with more than half of the galaxies having a pseudobulge. Towards lower and higher masses, the pseudobulge fraction declines (so that almost all pseudobulges are within $10^9<M_*<10^{11}$ M$_\odot$). These conclusions have been reproduced within models, be it directly through the pseudobulge fraction (\citealt{Shankar2013},~\citealt{Izquierdo-Villalba2019}) or indirectly through the mass fraction assembled through disc instabilities (\citealt{Parry2009},~\citealt{Tonini2016}). 

Model predictions on the effects of mergers can disagree due to different treatment of mergers. Broadly, these differences come from how models treat two important aspects of mergers: i) when galaxies merge and ii) what effect the merger has on the galaxies. Before considering how mergers affect galaxies, it is important to ensure that the frequency of mergers in a theoretical model matches that of the real Universe. Merger rates are usually used to quantify the statistics of mergers, as they are easy to calculate from galaxy formation models using merger trees  (\citealt{Maller2006},~\citealt{Stewart2009},~\citealt{Hopkins2010},~\citealt{Kaviraj2014},~\citealt{Lagos2018b},~\citealt{OLeary2020}).
In Paper I, we used the {\tt GALFORM} semi-analytical model (e.g. \citealt{Cole2000},~\citealt{Lacey2016}) to investigate the statistics of mergers in detail. We provided accurate merger rate, close pair fraction and merger timescale predictions up to $z=10$. 
We find that {\tt GALFORM} correctly predicts the frequency of mergers at least up to $z=3$. Beyond this redshift, it is hard to constrain any model due to inconsistent data from observations.

Here, our aims are twofold. The role of mergers in building up massive galaxies by $z=0$ is in fairly good agreement between models. However, there are disagreements at higher redshifts and for low-mass galaxies. There are also some disagreements on the impact of different merger types (major versus minor). To this end, we will investigate the predictions of {\tt GALFORM} regarding  the impact of mergers in the overall buildup of galaxy stellar mass. We will also quantify the relative importance of mergers and disc instabilities in building up galaxy spheroids. In a semi-analytical model it is possible to decompose their contributions, as well as that of the star formation bursts caused by both processes. A similar analysis was done by \cite{Parry2009}, for earlier versions of both the Durham semi-analytical model ({\tt GALFORM}; \citealt{Bower2006}) and the Munich semi-analytical model ({\tt L-GALAXIES}; e.g. \citealt{deLucia}). The version of {\tt GALFORM} we use in this paper (\citealt{Lacey2016}) is in better agreement with high-redshift observations, and treats mergers more accurately. Furthermore, we use the Planck Millennium simulation (\citealt{Baugh2019}), which has a better resolution both in terms of time (i.e. more outputs, important for mergers at high redshift) and mass (important for disc instabilities and minor mergers). These changes are likely to significantly affect predictions of {\tt GALFORM}. In this paper our focus is on a comprehensive comparison between {\tt GALFORM} and other models (including hydrodynamical simulations), as well as between our predictions and observations.



In Section~\ref{sec:Model} we introduce the {\tt GALFORM} galaxy formation model, and the Planck Millennium N-body simulation. We discuss how mergers, disc instabilities and the star formation in bursts caused by both are implemented in {\tt GALFORM}. In Section~\ref{sec:GalaxyGrowth} we compare the stellar mass growth rates due to star formation and mergers. We compare our predictions of the latter with observations. We also calculate the \textit{ex situ} mass fraction of galaxies, and compare our predictions with those of other models. In Section~\ref{sec:Spheroids} we explore the importance of various processes in the buildup of galaxy spheroids. We compare star formation rates in bursts caused by mergers and disc instabilities. Finally, we compare the contributions to the total spheroid mass of all channels of spheroid growth. In Section~\ref{sec:Conclusion} we summarise and conclude.




\section{Galaxy formation model}
\label{sec:Model}

We utilise the {\tt GALFORM} semi-analytical model of galaxy formation (\citealt{Cole2000}), which is implemented in the Planck Millennium N-body simulation (\citealt{Baugh2019}). {\tt GALFORM} models many physical processes that underpin galaxy formation, such as dark matter halo assembly, shock heating and radiative cooling of gas, the formation of galaxy discs, heating and ejection of gas due to supernovae and AGN feedback, galaxy mergers and disc instabilities and their effects on galaxy mass and morphology (\citealt{Cole2000},~\citealt{Baugh2006},~\citealt{Bower2006},~\citealt{GonzalezPerez2014},~\citealt{Lacey2016}). {\tt GALFORM} makes predictions that can be connected with observables, by following the chemical evolution of the gas and stars, dust emission/absorption and the stellar luminosity of galaxies. The latest version of {\tt GALFORM} with significant changes is presented in~\cite{Lacey2016}. \cite{Baugh2019} introduced a new galaxy merger scheme and a small recalibration of the parameters of the Lacey et~al. model for the Planck Millennium simulation. In this work we will not go into the details of how {\tt GALFORM} implements the physics of galaxy formation. These processes are explained in detail in~\cite{Lacey2016}. Instead we will focus on how {\tt GALFORM} models mergers, disc instabilities and their associated starbursts.

The Planck Millennium N-body simulation uses the cosmological parameters from the first year~\cite{PLANCK} data release\footnote{The cosmological parameters used are: $\Omega_\mathrm{M}=0.307$, $\Omega_\mathrm{\Lambda}=0.693$, $\Omega_\mathrm{b}=0.0483$, $h=0.677$, $\sigma_{8} = 0.8288$ and $n_{\rm s}=0.9611$.}. The simulation has a volume of (800 Mpc)$^3$ and uses 5040$^{3}$ particles. This number of particles corresponds to a resolution between those of the Millennium-I (\citealt{Springel2005}) and Millennium-II (\citealt{BoylanKolchin2009}) simulations. The particle mass is $\approx10^8 h^{-1}$ M$_\odot$, and the minimum halo mass corresponds to 20 particles. Halo merger trees are constructed using the 269 outputs of the Planck Millennium simulation. The halo and subhalo finder used is {\tt SUBFIND} (\citealt{Springel2001}). The {\tt DHALOS} algorithm (\citealt{Jiang2014b}) is used to construct halo merger trees.

\subsection{Galaxy mergers}
\label{sec:mergers}

As in all semi-analytical models, galaxy mergers require modelling choices to be made. This is because the resolution of large-volume, cosmological  N-body simulations such as the Planck Millennium is insufficient to follow the effects of dynamical friction for subhaloes with masses lower than a few $\times 10^{9}$ M$_\odot$. 
The subhalo merger scheme that we use was first implemented in the ~\cite{Baugh2019} version of {\tt GALFORM}, and it is motivated by the general approach outlined in \cite{Campbell2015}. The current merger scheme in {\tt GALFORM} is based on results of \cite{SimhaCole2017}. In short, this scheme assigns a merger time to subhaloes once they can no longer be resolved in the simulation. The fitting formula for the merger timescale was derived by matching the number of surviving haloes in the Millennium-I and Millennium-II simulations (see Paper I or \citealt{SimhaCole2017} for details). 

The calculation of the sizes of remnant galaxies after mergers are described in \cite{Lacey2016}, and in more detail in \cite{Cole2000}. Here, we are mainly interested in how mergers affect the mass evolution of spheroids, and how often they trigger starbursts. Morphological transformations and starbursts can be induced for a wide range of baryonic mass ratios $f_\mathrm{b}=M_\mathrm{b,sec}/M_\mathrm{b,pri}$, with $M_\mathrm{b,sec}$ and $M_\mathrm{b,pri}$ the baryonic masses of the secondary and primary galaxies, respectively. Increasing the mass ratio increases the probability of mergers inducing such events  (\citealt{Ellison2008},~\citealt{Scudder2012}). For simplicity, in {\tt GALFORM} this transition is assumed to be abrupt at some critical mass ratio. Pairs with $f_\mathrm{b}>f_\mathrm{ellip}$ are assumed to be major mergers. In major mergers, galaxy discs are destroyed, with all baryonic mass transferred to a newly formed spheroid. Mergers with $f_\mathrm{b}<f_\mathrm{ellip}$ are classified as minor: in this case, the stellar mass of the secondary is transferred to the (existing or newly formed) spheroid of the primary, while any cold gas in the secondary is transferred to the primary's disc. An additional threshold exists, $f_\mathrm{burst}<f_\mathrm{ellip}$, which separates minor mergers that trigger starbursts from those that do not. Both $f_\mathrm{ellip}$ and $f_\mathrm{burst}$ can in principle be determined from simulations. Such studies suggest $f_\mathrm{ellip}\approx0.3$ (e.g.~\citealt{Cox2009},~\citealt{Hopkins2009},~\citealt{Stewart2009b},~\citealt{Lotz2010}) and $f_\mathrm{burst}\approx0.1$ (e.g.~\citealt{Mihos1996},~\citealt{Birnboim2007},~\citealt{Cox2008}). In {\tt GALFORM}, both of these values are assumed to be free parameters. \cite{Lacey2016} used $f_\mathrm{ellip}=0.3$ and $f_\mathrm{burst}=0.05$. 

For the analysis here, we define major and minor mergers in a somewhat different way than described above, in order to simplify comparisons with other works. Instead of the baryonic mass ratio, we use the stellar mass ratio $\mu_*=M_\mathrm{*,sec}/M_\mathrm{*,pri}$. In addition, the limiting ratio between the two merger types is chosen to be $\mu_*=0.25$. Smooth accretion of satellites is defined as $\mu_*<0.1$. These definitions are used for consistency with other similar works in recent years. It should be noted that this definition of the limit between minor and major mergers is less physical than that used in the {\tt GALFORM} code (since dynamical disturbance to a galaxy should depend on the total masses involved, not only the mass in stars). As an example, defined in this way, minor mergers can sometimes trigger disc-to-spheroid transformations, while major mergers can sometimes not trigger them (both depending on gas fractions of the primary and secondary galaxies). 


\subsection{Disc instabilities}
\label{sec:instabilities}

Disc instabilities in {\tt GALFORM} are implemented in a fairly simple manner. We assume that a disc is gravitationally unstable to bar formation once it fulfills the instability criterion:
\begin{equation}
\frac{V_\mathrm{c}(r_\mathrm{disc})}{(1.68GM_\mathrm{disc}/r_\mathrm{disc})^{1/2}}<F_\mathrm{stab},
\label{eq:DiscInst}
\end{equation}
where $M_\mathrm{disc}$ is the baryonic mass of the disc, $r_\mathrm{disc}$ is its half-mass radius, $V_\mathrm{c}(r_\mathrm{disc})$ is the circular velocity at that radius, and $F_\mathrm{stab}$ is a parameter that controls which discs are unstable to bar formation (\citealt{Cole2000}). Discs are assumed to form a bar if the above condition is fulfilled. \cite{Efstathiou1982} estimated $F_\mathrm{stab}\approx1.1$ for stellar discs, while \cite{Christodoulou1995} found $F_\mathrm{stab}\approx0.9$ for gaseous discs. 

$F_\mathrm{stab}$ is allowed to vary in {\tt GALFORM} (see section 5.3 of \citealt{Lacey2016} for the effects of said variation). \cite{Lacey2016} chose the value $F_\mathrm{stab}=0.9$ in order to reproduce the present-day K-band luminosity function, and thus that value is used in the fiducial model. Note, however, that varying $F_\mathrm{stab}=0.9$ has an effect on many predictions of the model, such as the spheroid abundance at the present day and the UV luminosity function at high redshifts.

\cite{Izquierdo2022} studied the accuracy of the \cite{Efstathiou1982} criterion in the IllustrisTNG-50 (\citealt{Pillepich2019}) and IllustrisTNG-100 (\citealt{Nelson2019}) hydrodynamical cosmological simulations. They found that the instability criterion is accurate to within $\approx20\%$ in identifying discs that are stable against or unstable to bar formation. However, they used the criterion appropriate for stellar discs ($F_\mathrm{stab}=1.1$), so the accuracy of the instability criterion may be different for the value used in this paper ($F_\mathrm{stab}=0.9$).

Once a bar has formed, we assume that it buckles (\citealt{Combes1990},~\citealt{Friedli1993},~\citealt{Debattista2006}) and forms a spheroid. For simplicity, this process is assumed to be instantaneous (as soon as the disc becomes unstable). The resultant spheroid includes the stellar mass of the previous disc and spheroid, while the cold gas is fed to a starburst that is assumed to be triggered at the same time. The spheroid size is calculated in a manner similar to when a merger occurs (see~\citealt{Cole2000},~\citealt{Lacey2016}).

The \cite{Efstathiou1982} criterion may not be accurate for high-redshift gaseous discs (\citealt{Inoue2016}). These discs may instead undergo a different kind of instability, the so-called violent disc instability (see e.g. \citealt{Immeli2004}, \citealt{Dekel2009}, ~\citealt{Elmegreen2008},~\citealt{Guo2015})). In this instability mode, the disc forms massive clumps that may eventually migrate to the centre of the galaxy, where they may form a bulge or pseudobulge. However, it is not clear how prevalent this mode of instability is in the real Universe, nor how reliable our understanding of these instabilities is from simulations, given their dependence on the modeling details, such as the supernova feedback scheme (e.g. \citealt{Cacciato2012}). In {\tt GALFORM} this mode of instabilities is currently not included, but we may include it in the future. The secular evolution of discs (e.g.~\citealt{Genzel2008},~\citealt{Sellwood2014}) is also not included in {\tt GALFORM}. Instead, entire galaxy discs are assumed to evolve into spheroids once the instability criterion is fulfilled. While this may be a somewhat extreme implementation, discs generally regrow fairly quickly in {\tt GALFORM}.


\subsection{Starbursts}
\label{sec:starbursts}

Observationally, starbursts are assumed to be going on within a galaxy if its SFR or specific SFR (sSFR hereafter) is above some threshold value. In {\tt GALFORM}, starbursts are instead assumed to be triggered by dynamical disturbances, which are in our case galaxy mergers and disc instabilities. Starbursts are triggered in {\tt GALFORM} for all disc instabilities, as well as minor and major mergers mergers with a baryonic mass ratio $f_\mathrm{b}>0.05$. The SFR in a burst is given by
\begin{equation}
\psi_\mathrm{burst}=\frac{M_\mathrm{cold}}{\tau_\mathrm{burst}},
\label{eq:BurstSFR}
\end{equation}
where $M_\mathrm{cold}$ is the cold gas mass that includes the gas transferred to the spheroid as part of the triggering event, as well as any cold gas remaining from a previous starburst. $\tau_\mathrm{burst}$ is a starburst timescale given by:
\begin{equation}
\tau_\mathrm{burst}=\mathrm{max}(f_\mathrm{dyn}\tau_\mathrm{dyn,sph},\tau_\mathrm{burst,min}).
\label{eq:BurstTimeScale}
\end{equation}
In this definition, $\tau_\mathrm{burst,min}$ is a floor value adopted for the starburst timescale, $\tau_\mathrm{dyn,sph}=r_\mathrm{sph}/V_\mathrm{c}(r_\mathrm{sph})$ is the dynamical timescale of the spheroid, and $f_\mathrm{dyn}$ is a free parameter. The scaling between the starburst and spheroid dynamical timescales is suggested by observations (\citealt{Kennicutt1998},~\citealt{Biegel2008}), and is parametrised by $f_\mathrm{dyn}$. The values used for $f_\mathrm{dyn}$ are fairly large; \cite{Lacey2016} proposed $f_\mathrm{dyn}=20$. Note that $M_\mathrm{cold}$ in Eqn.~(\ref{eq:BurstSFR}) represents the current amount of cold gas remaining in a starburst, so it decreases with time exponentially. The starburst is assumed to stop after 3 e-folding times. Any galaxy with a non-zero starburst SFR is assumed to have an ongoing starburst, regardless of when it was initiated. We use the above definition of a starburst throughout the paper (instead of the observationally-motivated one, which uses a threshold value of the SFR or sSFR).




\section{Galaxy growth: star formation vs. mergers}
\label{sec:GalaxyGrowth}

Star formation and mergers are the two main channels of galaxy stellar mass growth. In this section we compare the contribution of each channel to the growth of galaxies, through the \textit{in situ} mass growth rate and the \textit{ex situ} fraction. We investigate the dependencies of both quantities on stellar mass and redshift, and also calculate the contributions of various merger types to the total mass gained through mergers.

\subsection{Stellar mass growth rates}
\label{sec:GrowthRates}

The relative roles of star formation and mergers in the stellar evolution of galaxies can be compared by calculating the growth rate through each channel. We divide by the current stellar mass to obtain the specific stellar mass growth rate (sSFR), $\dot{M}_*/M_*$. We do this to facilitate an easy comparison between galaxies of different masses. Note that by 'star formation rates', we are referring to stellar mass growth rates due to star formation. This can generally be different from the total star formation rate, since some portion of the stellar mass is immediately recycled (see \citealt{Lacey2016} for details). For the merger growth rate we include all mergers, including those with low mass ratios, since the mass growth rate due to mergers does not diverge (unlike the merger rate).

\begin{figure*}
\includegraphics[width=0.99\textwidth, trim = 0 15 0 0]{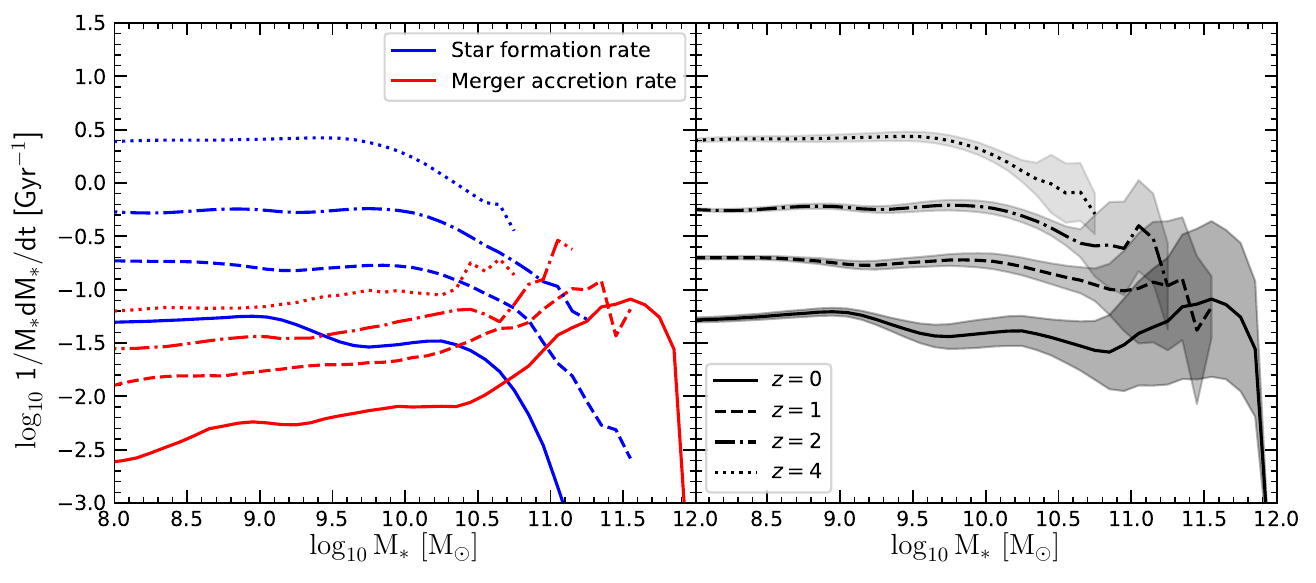}
\caption{Specific stellar mass growth rates of galaxies through \textit{in situ} star formation (blue lines) and merger-induced accretion of stars formed \textit{ex situ} (red lines). Line types represent different redshifts, as per the legend in the right-hand panel. Lines are discontinued at the highest mass at which galaxies or mergers are present in the sample. \textit{Left:} Growth rates split by channel. \textit{Right:} Total growth rate. Shaded areas represent 16 to 84 percentile ranges.}
\label{fig:sSMGR}
\end{figure*}

In Fig.~\ref{fig:sSMGR} we show the predicted growth rates due to both channels. Star formation clearly dominates for most of the stellar mass range, and at all redshifts. At $z=0$, mergers begin to dominate the growth of galaxies more massive than $M_*=10^{10.7}$ M$_\odot$. This transitional mass increases at higher redshifts, reaching $M_*=10^{11}$ M$_\odot$ at $z=2$. At even higher redshifts ($z=4$) it would appear that star formation dominates for all galaxies, but we note that, due to the evolution of the stellar mass function, our sample does not include any $M_*>10^{11}$ M$_\odot$ galaxies at this redshift. From the right-hand panel we see that the total (specific) growth rate is constant as a function of stellar mass, except for the very high-mass end of the existing population at a given redshift. This implies that the evolution of the stellar mass of galaxies with $M_*<10^{10.5}$ M$_\odot$ is self-similar.

The general picture presented in Fig.~\ref{fig:sSMGR} is not surprising. Similar conclusions for the relative roles of the two growth channels have been found in previous semi-analytical models (e.g. \citealt{Guo2008}), semi-empirical models (e.g. \citealt{Moster2013}) and hydrodynamical simulations (e.g. \citealt{RodrGomez2016}). Unlike most of these studies, however, we predict a plateau and subsequent fall in the merger growth rate at very high masses ($M_*=10^{11.5}$ M$_\odot$ at $z=0$). This feature can only be seen with a sufficiently large sample size: the $z=0$ merger growth rate in Fig.~\ref{fig:sSMGR} represents $\approx16$ million events. 

In Paper I, this decline at very high masses was easier to see at higher redshifts because the quantity of interest was the merger rate. The merger rate is usually split into major ($\mu_* \in[0.25,$ $1]$) and minor ($\mu_* \in[0.1,$ $0.25]$) merger rates. We found clear evidence of a fall in the former, while the latter showed only some signs of a decline, and only at higher masses. We argued that this decline is the result of a suppressed number of galaxies seen in the galaxy stellar mass function (GSMF) beyond the `knee'. This affects mergers by reducing the number of available companions. For major mergers this decline in the number of companions can be seen at a lower host stellar mass: $M_*=10^{11.2}$ M$_\odot$ (at $z=0$). For minor mergers this corresponds to $M_*=10^{11.7}$ M$_\odot$. If accretion ($\mu_* \in[0,$ $0.1]$) were included, there might be no maximum and decline, as this would include low-mass galaxies (at masses below the break in the GSMF). As seen in Fig.~\ref{fig:sSMGR}, the combined effect of all merger types is to create a peak and then a decline in the merger growth rate at $M_*=10^{11.5}$ M$_\odot$. At higher redshifts we do not see evidence of the same decline, but we suspect this is the result of an insufficient sample size at very high masses. 

We note that this decline was not found in Illustris (\citealt{RodrGomez2015}), nor in observations by \cite{Robotham2014}, who studied close pair fractions as a function of stellar mass in the GAMA survey. Instead, \cite{Robotham2014} find that the merger growth rate sharply increases above $M_*=10^{11.3}$ M$_\odot$. In obtaining this and their other results, they used the merger timescale derived by \cite{KitzWhite2008}. In Paper I we found a stronger dependence of the merger timescale on stellar mass than \cite{KitzWhite2008}, but this is unlikely to account for the stark difference between the growth rates we predict and those found by \cite{Robotham2014} (strong decline vs. strong increase at very high masses).

Here we will not compare our predicted SFR with observations, since this has already been done for GALFORM in \cite{Lacey2016} and \cite{Cowley}. The same is not true of the merger growth rate. Massive galaxies in the local Universe are the best candidates for such a comparison, as this is the regime in which most measurements have been carried out. 
We compare our results with \cite{McIntosh2008}, \cite{LopezSanjuan2012} and \cite{Ferreras2014}, all of whom studied the stellar mass growth rates of galaxies with $M_*>10^{11}$ M$_\odot$, due to mergers. They found specific stellar mass growth rates $\dot{M}_*/M_*$ equal to $6\pm3$, $1-9$ and $8\pm2$ in units of \% of mass added per Gyr, respectively. This is comparable to our values for $M_*>10^{11}$ M$_\odot$ and $z<1$ of $2-10$ \%Gyr$^{-1}$.

\subsection{\textit{Ex situ} fraction}
\label{sec:ExSitu}

In recent years it has become popular to compare the impact of star formation and mergers using the fraction of stellar mass formed \textit{ex situ}. This is defined simply as the fraction of stellar content that a given galaxy accreted through mergers, with the remainder attributed to \textit{in situ} star formation. This is a useful quantity for comparisons since it is dimensionless and has a clear physical meaning.

In principle, one avenue to calculate the \textit{ex situ} fraction is by using the growth rates outlined above. The stellar mass history of galaxies can be calculated from the total stellar mass growth rate, which can then be used alongside the merger growth rate to calculate the \textit{ex situ} mass history. However, this approach has some deficiencies, namely: i) it requires using mean growth rates, so the scatter in the population is lost, ii) errors propagate during the integration, iii) growth rates for very massive galaxies at a given redshift are uncertain (not smooth) leading to the same issue in the calculated \textit{ex situ} fraction.

Instead of the above approach, we choose to calculate the \textit{ex situ} fraction directly using all of the information available from the model. This is done by using galaxy merger trees: for every galaxy we add up the stellar mass of all galaxies merging onto the main or most massive branch of the merger tree. This process is done by navigating along merger trees from the redshift of interes, $z$, backwards to the first output. The stellar mass obtained through this algorithm, which we denote by $M_\mathrm{*,ex}$, represents the \textit{ex situ} mass of a given galaxy of mass $M_*$ at redshift $z$.

The \textit{ex situ} fraction of galaxies can be estimated from observations of diffuse stellar haloes, which are thought to be built primarily by tidal stripping of infalling satellite galaxies. The stellar mass of the halo can be estimated by means of a photometric decomposition (e.g. \citealt{Spavone2017}). In this method, multiple components (with different Sérsic indices) are assumed to make up the surface brightness profile of a given galaxy. The outermost component, i.e. the stellar halo, is then associated with mergers, and the inner one with \textit{in situ} processes (such as pseudobulge formation through disc instabilities, e.g. \citealt{Fisher2011}). There are a few caveats associated with this method, namely: i) some of the stellar halo could be associated with tidal streams of stars pulled out from the central galaxy during mergers and ii) some of the inner mass is likely associated with mergers, since mergers do not only build up the stellar halo. We compare our results with estimates obtained through this method despite these complications$-$they are the first and best estimates available so far. In particular, we compare with the compilation of observational estimates presented in \cite{Spavone2021}. This compilation includes observational estimates of stellar halo mass fractions from \cite{Seigar2007}, \cite{Iodice2016}, \cite{Spavone2017}, \cite{Spavone2018}, \cite{Cattapan2019}, \cite{Spavone2020}, \cite{Iodice2020} and \cite{Spavone2021}. We do not include galaxies with $M_*<10^{11}$ $\mathrm{M}_\odot$ from the sample of \cite{Spavone2017}, since that sample contains only galaxies from the Fornax cluster, and it is likely that results for low-mass galaxies from such a sample are not representative of the whole population. 

We also compare our results with other theoretical predictions. Results on the \textit{ex situ} fraction are available from the {\tt EMERGE} semi-empirical model (\citealt{Moster2018}), the EAGLE (\citealt{Schaye2015}), Illustris (\citealt{Vogelsberger2014}), IllustrisTNG (\citealt{Pillepich2018}) and Horizon AGN (\citealt{Dubois2014}) hydrodynamical simulations, as well as other semi-analytical models (\citealt{Henriques2015}, \citealt{LeeYi2017}). The results on the \textit{ex situ} fraction for the semi-empirical model and hydrodynamical simulations are given in \cite{Moster2020}, \cite{Davison}, \cite{RodrGomez2016}, \cite{Tacchella2019} and \cite{Dubois2016}, respectively. The results from the \cite{Henriques2015} semi-analytical model are presented in \cite{Henriques2019}.

\begin{figure*}
\includegraphics[width=0.99\textwidth, trim = 0 15 0 0]{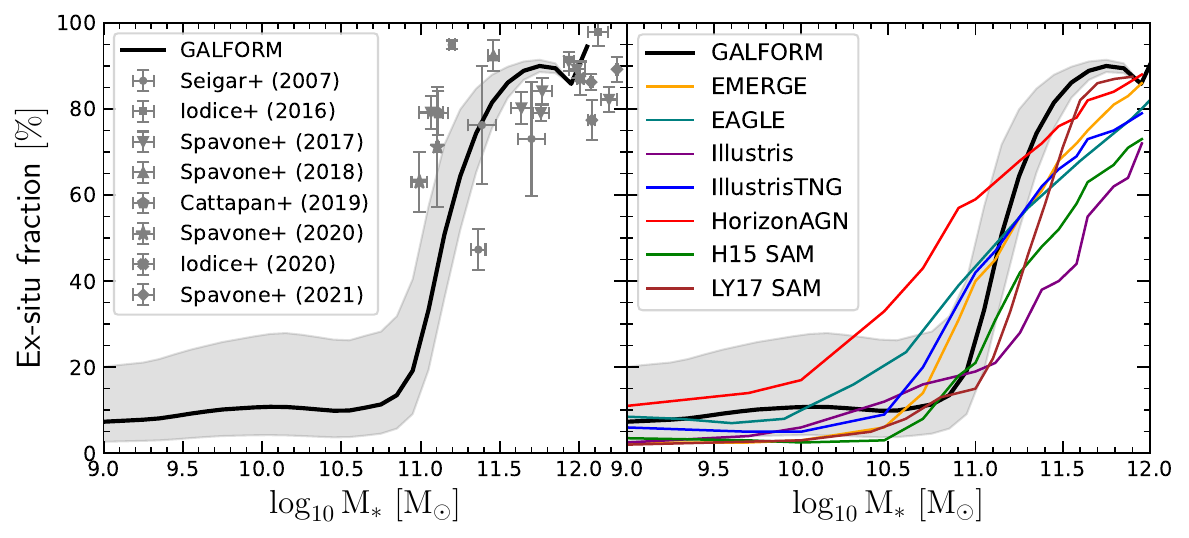}
\caption{\textit{Ex situ} fraction of galaxies in GALFORM (black line), in comparison with observational estimates of stellar halo mass fractions (\textit{left-hand panel}) and a suite of other models (coloured lines in the \textit{right-hand panel}, as per the legend). Results are presented for the local universe ($z=0$). The shaded area represents the 16 to 84 percentile range of the \textit{ex situ} fraction in GALFORM.}
\label{fig:ExSituModels}
\end{figure*}

Fig.~\ref{fig:ExSituModels} shows the predicted \textit{ex situ} fraction from {\tt GALFORM} as a function of stellar mass for galaxies at $z=0$. The \textit{ex situ} fraction is low ($7-12\%$) up to a high stellar mass $M_*\sim 10^{11}$ M$_\odot$, beyond which it rises sharply. This finding is expected given the discussion in the previous subsection about growth rates through star formation and mergers. It also coincides with the mass regime in which the majority of galaxies transition from being disc-dominated to spheroid-dominated (e.g. \citealt{Poggianti2009}, \citealt{Brennan2015}).

The scatter around the mean \textit{ex situ} fraction in {\tt GALFORM} is fairly significant for all but the most massive galaxies. In fact, from Fig.~\ref{fig:ExSituModels} we conclude that the scatter falls with increasing stellar mass. This is because massive galaxies are constrained to grow through a single mechanism: mergers. It is thus not surprising that such galaxies exhibit little variation in their \textit{ex situ} fraction. On the other hand, galaxies of stellar mass $M_*=10^{9}$ M$_\odot$ have a mean \textit{ex situ} fraction $f_\mathrm{ex}=0.07$, but the 16 to 84 percentile range corresponds to $f_\mathrm{ex}=0.03-0.2$. This means that mergers are a much more stochastic process for low-mass galaxies. Some of them never experience a merger, while others are significantly affected either by many smaller mergers or a single major merger. This behaviour in the scatter is not unexpected. Massive galaxies in clusters are in a position to merge more often, while lower-mass galaxies are likely to be field or group galaxies, and can thus merge only if their relative separation and velocity allow it. 

In Fig. \ref{fig:ExSituModels} we also compare the findings from {\tt GALFORM} with observational estimates of stellar halo mass fractions (left-hand panel) and results on \textit{ex situ} fractions from other models (right-hand panel). Our predictions are in good agreement with observations: the most massive galaxies have \textit{ex situ} in the range $80-100\%$, and the transition from the regime of low \textit{ex situ} fractions to that where galaxies have $>50\%$ is around $M_*=10^{11}$ $\mathrm{M}_\odot$. However, the the number of observed galaxies with estimated \textit{ex situ} fractions is too low to offer strong constraints on the details of the model (e.g. the shape of the curve, the exact transition mass or the maximal \textit{ex situ} fraction reached by the most massive galaxies). 

From the right-hand panel we see that {\tt GALFORM} differs from most other models in that it predicts a fairly sharp increase in the \textit{ex situ} fraction. This increase begins at $M_*\approx10^{10.9}$ M$_\odot$, where we have $f_\mathrm{ex}=0.1$, and values of $f_\mathrm{ex}=0.9$ are reached by $M_*\approx10^{11.5}$ M$_\odot$. This sharp increase is possibly related to the fact that in {\tt GALFORM}, AGN feedback is turned on sharply, with gas cooling completely suppressed in massive haloes. Other models, with the exception of \cite{LeeYi2017}, predict a more protracted rise, beginning from anywhere between $M_*=10^{10}$ M$_\odot$ and $M_*=10^{11}$ M$_\odot$. The behaviour predicted by {\tt GALFORM} is similar to \cite{LeeYi2017}, although the sharp increase in that model begins $\approx0.1-0.2$ dex higher in stellar mass (at $M_*\approx10^{11.1}$ M$_\odot$). The eventual values of the \textit{ex situ} fraction reached by $M_*=10^{12}$ M$_\odot$ vary between $f_\mathrm{ex}=0.7$ to $f_\mathrm{ex}=0.9$ from model to model. The latter value is found in {\tt EMERGE}, Horizon AGN and \cite{LeeYi2017}, alongside {\tt GALFORM}.

The low-mass ($M_*<10^{10.5}$ M$_\odot$) behaviour predicted by {\tt GALFORM} also differs somewhat from all other models. The other theoretical predictions, with the exception of Horizon AGN, suggest fairly low and decreasing \textit{ex situ} fractions as we go to lower masses, with values below $f_\mathrm{ex}=0.05$. On the other hand, {\tt GALFORM} predicts fairly constant values between $f_\mathrm{ex}=0.07-0.12$. There is also a slight `dip' around $M_* \approx10^{10.6}$ M$_\odot$, implying that there is a complex interplay between mergers and other processes. As we show in Section~\ref{sec:Spheroids}, this dip is due to an increased significance of disc instabilities at these masses.


\subsection{Redshift dependence}
\label{sec:ExSituRedshift}

\begin{figure}
\includegraphics[width=0.99\columnwidth, trim = 0 15 0 0]{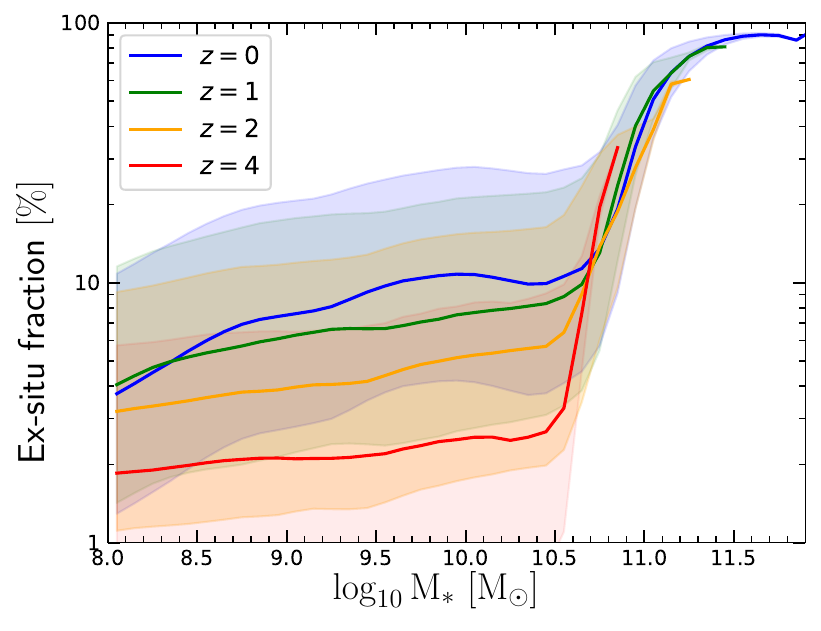}
\caption{\textit{Ex situ} fraction of galaxies at different redshifts, as per the legend. Lines are discontinued at the mass for which our sample contains no galaxies, or where all galaxies in the sample have an exactly-zero \textit{ex situ} stellar mass. Shaded areas represent the 1-$\sigma$ scatter around the mean value.}
\label{fig:ExSituRedshift}
\end{figure}

In the previous subsection we focused on the stellar mass dependence of the \textit{ex situ} mass fraction at $z=0$. We now turn to higher redshifts. In Fig.~\ref{fig:ExSituRedshift} we show the {\tt GALFORM} predictions for the same quantity up to $z=4$. The transition to high $f_\mathrm{ex}$ values occurs at the same mass for all redshifts: $M_*\approx10^{10.7}$ M$_\odot$. Furthermore, the shape of the transition is the same at all redshifts. The behaviour of this transition is consistent with the sudden turning on of AGN feedback (see \citealt{Lacey2016} for details of the AGN feedback as implemented in {\tt GALFORM}). The characteristic halo mass by which AGN feedback is turned on is $M_\mathrm{h}=10^{12.5}$ M$_\odot$, independent of redshift (\citealt{Mitchell2016}, \citealt{Baugh2019}). Coupled with negligible evolution in the knee of the stellar to halo mass relation, at least in {\tt GALFORM} (\citealt{Mitchell2016}),  the turnover in the \textit{ex situ} fraction can be attributed to AGN feedback. For lower mass galaxies ($M_*<10^{10.5}$ M$_\odot$), we can see a clear increase in the importance of mergers at later times. In this regime, the \textit{ex situ} fractions are 5 times higher in the local Universe than at $z=4$.

For high-mass galaxies, the constant \textit{ex situ} fraction in {\tt GALFORM} indicates that these galaxies grew in a similar way. This is a result of quenching at the same halo (and stellar) mass, and the fact that quenching is sudden. In order to reach some large stellar mass in this simple picture, where effectively all \textit{in situ} star formation is stopped around $M_*=10^{10.5}$ $\mathrm{M}_\odot$, all galaxies more massive than that by the same amount should have a similar \textit{ex situ} fraction, regardless of redshift. This interpretation is in line with our finding that the \textit{ex situ} fraction is $f_\mathrm{ex}\approx90\%$ by $M_*=10^{11.5}$ $\mathrm{M}_\odot$. The high-mass behaviour in hydrodynamical simulations (a falling \textit{ex situ} fraction with redshift, at the same stellar mass) indicates that massive galaxies are less quenched at higher redshifts.

The redshift dependence of \textit{ex situ} fraction is especially interesting to compare with other models, as most models are calibrated to reproduce the $z=0$ stellar mass function. For this reason it is not surprising that they produce a similar picture (Fig.~\ref{fig:ExSituModels}). Unfortunately, many different mass selections are used when studying the dependence of $f_\mathrm{ex}$ on redshift, so it is difficult to compare results quantitatively. Instead, we will make a qualitative comparison with other models.

The Illustris and Illustris TNG simulations (\citealt{RodrGomez2016}, \citealt{Gupta2020}), the EAGLE simulation (\citealt{Schaye2015}, \citealt{Qu2017}) and the {\tt EMERGE} semi-empirical model all predict an \textit{ex situ} fraction that remains constant or rises with redshift for low-mass galaxies. On the other hand, for high-mass galaxies they predict a fall with redshift. This is different to our predictions from {\tt GALFORM}, on both counts. We find a constant \textit{ex situ} fraction at high masses, and a falling (with redshift) fraction at low masses.


\subsection{Contribution of different merger types to mass growth}
\label{sec:MergerTypes}

Here we will look at the contribution of various merger types to the total \textit{ex situ} mass accreted by galaxies. We classify mergers into three categoeies: i) major mergers, $\mu_*\in[0.25,1]$; ii) minor mergers, $\mu_*\in[0.1,0.25]$ and iii) accretion, $\mu_*<0.1$. We quantify the contribution to stellar mass growth of each channel through the ratio $f_\mathrm{merg,i}/f_\mathrm{ex}=M_\mathrm{*,merg,i}/M_\mathrm{*,ex}$, where $M_\mathrm{*,merg,i}$ is the stellar mass attributed to the i-th channel, and $f_\mathrm{merg,i}$ the same quantity divided by the total stellar mass.

\begin{figure}
\includegraphics[width=0.99\columnwidth, trim = 0 15 0 0]{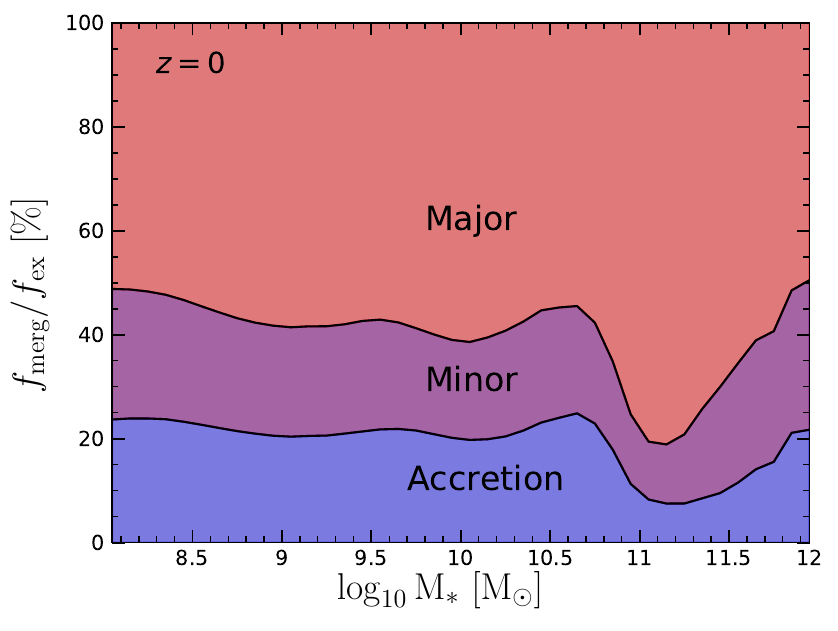}
\caption{Fraction of total \textit{ex situ} mass contributed by different growth channels, for the local Universe ($z=0$). These are defined as: accretion; $\mu_* \in [0,0.1]$, minor mergers; $\mu_* \in [0.1,0.25]$, major mergers; $\mu_* \in [0.25,1]$. The proportion of total merger mass attributed to each channel is represented by differently coloured areas, as labeled on the plot.}
\label{fig:ExSituTypes}
\end{figure}

Fig.~\ref{fig:ExSituTypes} shows the contribution of each channel of merging to the total \textit{ex situ} mass of galaxies at $z=0$, as a function of stellar mass. We find that at most masses major mergers contribute roughly $50-60\%$, and minor mergers about the same as accretion, at $20-25\%$. We have checked these dependencies out to high redshifts, and they remain similar. Major mergers increase their contribution to $80\%$ by $M_*=10^{11}$ M$_\odot$, and this increase coincides with the sharp rise in the total \textit{ex situ} fraction of galaxies. Beyond this mass, major mergers begin to decline again. This is due to the fall in the number count of galaxies seen in the GSMF, as already discussed in Section~\ref{sec:GrowthRates}.

The contribution of different merger types to galaxy mass growth is another interesting area of disagreement among models. The Illustris simulation predicts roughly the same contributions as we do, but without a rise and subsequent fall for major mergers at $M_*>10^{10.5}$ M$_\odot$ (\citealt{RodrGomez2016}). The rise and fall that we predict is seen in one form or another in most other models. What differs is the exact shape of the curves. Furthermore, some models predict overall contributions that do not match ours even roughly. {\tt EMERGE} predicts a contribution of $40\%$ each through major mergers and accretion at $M_*<10^{10.5}$ M$_\odot$, while for $M_*>10^{10.5}$ M$_\odot$ major mergers contribute up to $90\%$ (\citealt{OLeary2020}), but then decline towards the highest masses to $80\%$. The semi-empirical model of \cite{Hopkins2010} predicts a similar decline, with minor mergers also contributing $30\%$ at lower masses. These differences may be due to different mass resolutions used in the various simulations.


In the IllustrisTNG simulation, minor mergers and accretion dominate over major mergers for $M_*<10^{10.5}$ M$_\odot$, while the reverse is true for $M_*>10^{10.5}$ M$_\odot$ (\citealt{Tacchella2019}). \cite{Cattaneo2011} studied mergers in a semi-analytical model, and investigated the contribution of major mergers using two different definitions: $\mu_* \in[1/5,1]$ and $\mu_* \in[1/3,1]$. We take the mean of these two values to compare with our choice of $\mu_{*,\mathrm{lim}}=0.25$. With this definition, their model predicts a major merger contribution of only $20\%$ at $M_*=10^{10}$ M$_\odot$, and $60\%$ for $M_*>10^{11}$ M$_\odot$. Their high-mass predictions are in good agreement with ours, but the low-mass ones are not.

The relative contributions of merger types can also be compared with observations, but usually only for the most massive systems. Observational measurements suggest that the present-day growth of $M_*=10^{11}$ M$_\odot$ galaxies is mainly through major mergers (\citealt{McIntosh2008}, \citealt{Ferreras2014}). This is in good agreement with our results. With a somewhat different selection, $M_*>10^{11}$ M$_\odot$, \cite{LopezSanjuan2012} measured that major and minor mergers contribute $75\%$ in combination, while accretion contributes $25\%$. This is qualitatively in good agreement with our prediction of a waning major merger contribution beyond $M_*=10^{11}$ M$_\odot$. \cite{Tal2012} studied luminous red galaxies (LRGs) and found that interactions with satellites are mostly restricted to minor mergers, in agreement with the trends we predict. 


The \textit{ex situ} fraction depends on both stellar mass and redshift: $f_\mathrm{ex}=f_\mathrm{ex}(M_*,z)$. By using the galaxy stellar mass function and integrating, we can calculate the mean \textit{ex situ} contribution to the stellar mass of the universe. We use the following formula:
\begin{equation}
f_\mathrm{ex,global}(z)=\frac{\int_{0}^{\infty} f_\mathrm{ex}(M_*,z)M_*\Phi(M_*,z)\hspace{0.3mm}\mathrm{d}M_*}{\int_{0}^{\infty} M_*\Phi(M_*,z)\hspace{0.3mm}\mathrm{d}M_*},
\label{eq:MeanExSitu}
\end{equation}
with $\Phi(M_*,z)$ the galaxy stellar mass function. The lower bound in these integrals is set to $0$ since they converge (unlike the integrals of the GSMF itself). In addition, the number of low-mass galaxies begins to decrease below $M_*=10^7$ M$_\odot$, although this is an artefact of numerical resolution. Note that we have inserted a factor $M_*$ in both integrals, so that the denominator is the total stellar mass density of the Universe, while the numerator is the total \textit{ex situ} stellar mass density. 

\begin{figure}
\includegraphics[width=0.99\columnwidth, trim = 0 15 0 0]{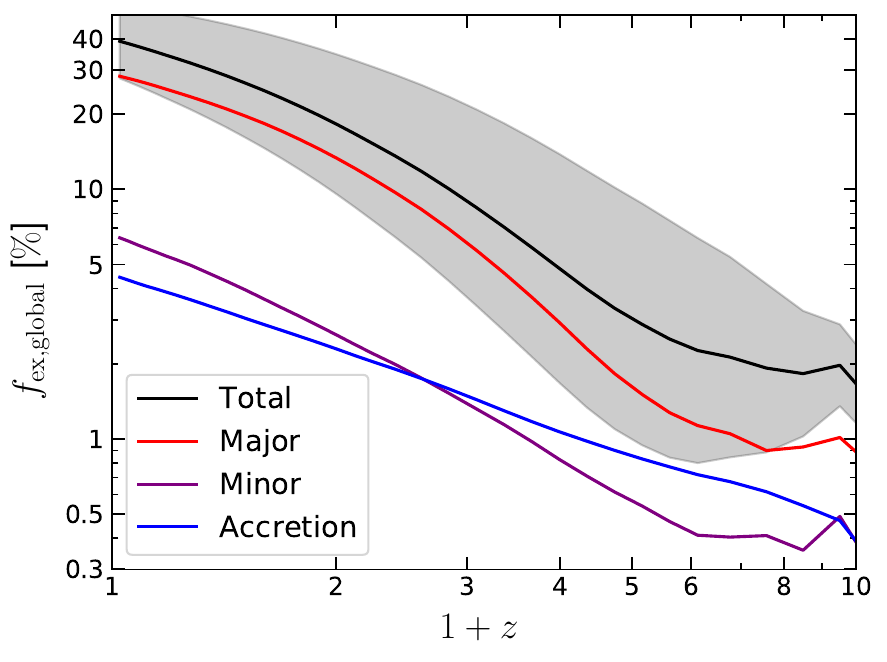}
\caption{\textit{Ex situ} fraction of the total stellar mass of the universe, as a function of redshift. The black line (mean) and associated shaded area (16 to 84 percentile range) represent the total \textit{ex situ} fraction, while coloured lines represent each channel of merger growth, as per the legend. These channels are defined as: accretion; $\mu_* \in [0,0.1]$, minor mergers; $\mu_* \in [0.1,0.25]$, major mergers; $\mu_* \in [0.25,1]$.}
\label{fig:ExSituGlobal}
\end{figure}

In Fig.~\ref{fig:ExSituGlobal} we show the evolution of the global \textit{ex situ} fraction, as well as the contributions of the different merger channels. 40\% of the stars in the local universe reside in systems in which they did not form. The same is true for {10\%, 3\%, 2\%} of stars at $z = {2,4,8}$, respectively. Our predictions are qualitatively similar to those from the Illustris simulation \citep{RodrGomez2016}, though somewhat different in detail:  for Illustris the figures reported are: {$30\%$ at  $z=0$, $16\%$ at $z=2$, $12\%$ at $z=4$}. In Illustris  mergers clearly have a larger influence overall at higher redshifts. The contribution of major mergers dominates the global \textit{ex situ} fraction at all redshifts. At very high redshifts ($z>4$), the contribution of accretion is somewhat closer to that of major mergers. Minor mergers became more important at $z\approx2$. The present day \textit{ex situ} mass fraction ($40\%$) is built through: $29\%$ major mergers, $6.5\%$ minor mergers and $4.5\%$ accretion.

\section{Growth of galaxy spheroids}
\label{sec:Spheroids}

Galaxy spheroids are created or enlarged after mergers and disc instabilities. These events cause the transformation of discs (or whole galaxies) into spheroids (a classical spheroid if from a merger, or a pseudobulge if from a disc instability\footnote{Note that some simulations imply that pseudobulges can also be created through mergers (\citealt{Keselman2012}, \citealt{Wang2015}).}), but they also trigger bursts of star formation. These bursts of star formation also contribute to the stellar mass growth of spheroids.

\subsection{Spheroid mass fractions}

\begin{figure*}
\includegraphics[width=0.99\textwidth, trim = 0 15 0 0]{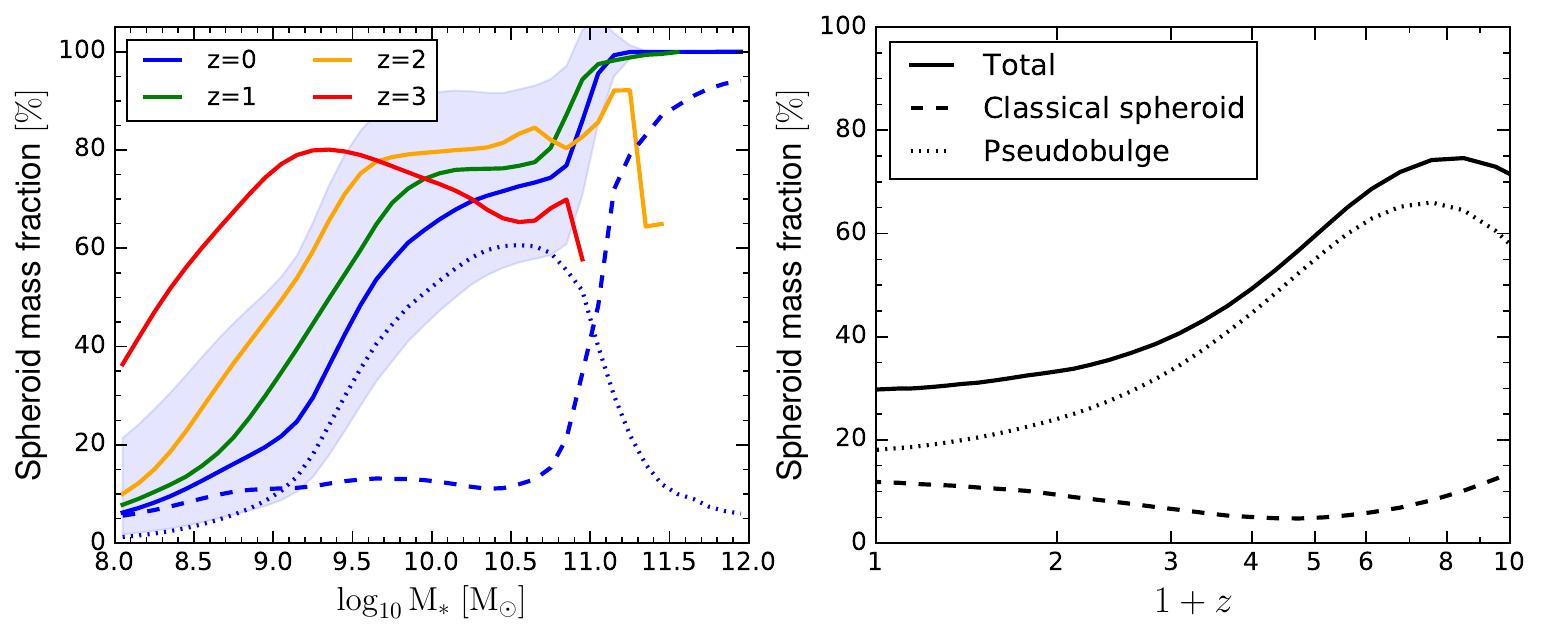}
\caption{The spheroid mass fraction (fraction of stellar mass in spheroids relative to the total stellar mass of a galaxy) as a function of stellar mass at several redshifts (\textit{left}), and the global spheroid mass fraction as a function of redshift (\textit{right}). We also show the contributions of mergers and disc instabilities to the total spheroid mass fraction as the dashed and dotted lines (classical spheroids and pseudobulges, respectively), for the $z=0$ line in the left-hand panel and the entire population in the right-hand panel.}
\label{fig:SphMassFrac}
\end{figure*}

We begin our analysis of spheroid growth in {\tt GALFORM} by quantifying how widespread spheroids are, as a function of both stellar mass and redshift. For this purpose we focus on the spheroid mass fraction, i.e. the spheroid-to-total mass ratio $M_\mathrm{*,sph}/M_\mathrm{*,total}$ (or $B/T$ as it is commonly referred to, standing for bulge/total). The left-hand panel of Fig. \ref{fig:SphMassFrac} shows the spheroid mass fraction as a function of stellar mass at several redshifts. The spheroid mass fraction grows with stellar mass at almost all redshifts, indicating the progressively stronger effects of disc instabilities and mergers. 

For the $z=0$ curve in the left-hand panel of Fig. \ref{fig:SphMassFrac}, we show the contributions of mergers and disc instabilities to the spheroid mass fraction separately (with dashed and dotted lines, respectively). For this purpose, and for the remainder of the paper, we assume that classical spheroids and disc instabilities are built not only directly through these two processes, but also through the starbursts that they trigger. From this decomposition we see that the complex shape of the curve is a result of disc instabilities dominating between $M_*=10^9$ $\mathrm{M}_\odot$ and $M_*=10^{11}$ $\mathrm{M}_\odot$, and mergers dominating outside this range. At higher stellar masses the spheroid mass fraction is effectively $100\%$, while it is $5-10\%$ for $M_*<10^9$ $\mathrm{M}_\odot$. 

From the left-hand panel of Fig. \ref{fig:SphMassFrac} we see that at almost all masses, the spheroid mass fraction grows with redshift. This is mainly due to the increasing effect of disc instabilities. In the right-hand panel we show the global spheroid mass fraction as a function of redshift (calculated in a an analogous way to Eqn. \ref{eq:MeanExSitu}), as well as the contributions to the total mass fraction from classical spheroids and pseudobulges. We see that the spheroid mass fraction is very high ($\approx80\%$) at high redshifts, $z>6$. At lower redshifts the spheroid mass fraction falls to a roughly constant $30\%$ by $z=2$. At all redshifts the global spheroid mass fraction is dominated by pseudobulges, but mergers are progressively more important at lower redshifts, as seen in the higher classical spheroid mass fraction. At $z=0$, pseudobulges make up around two thirds of all spheroid mass, with the rest in classical spheroids.

\subsection{Spheroid growth due to mergers} 

\begin{figure}
\includegraphics[width=0.99\columnwidth, trim = 0 15 0 0]{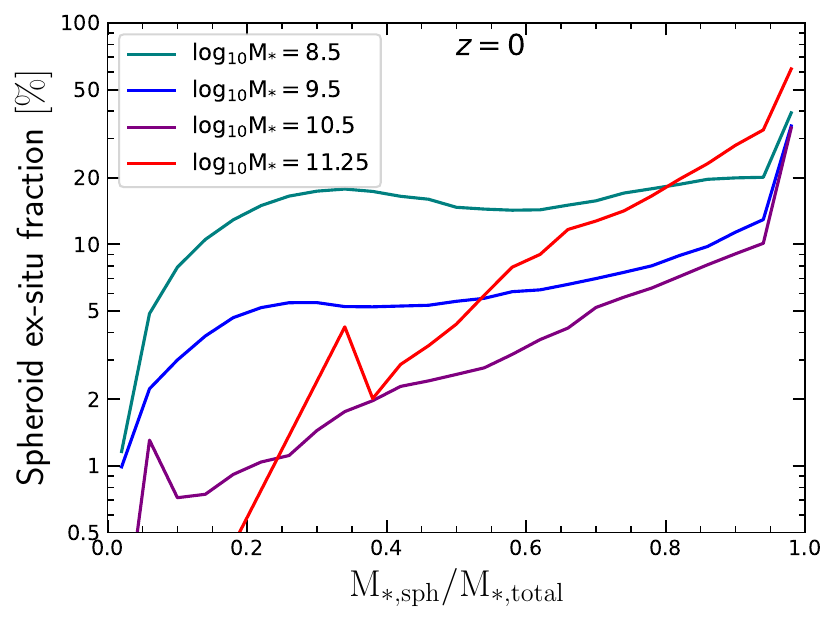}
\caption{Mean \textit{ex situ} fraction of galaxy spheroids as a function of spheroid mass fraction, for the local universe ($z=0$). Line colours represent different galaxy masses, as indicated by the  legend.}
\label{fig:ExSituSph}
\end{figure}

We will now study the growth of spheroids in {\tt GALFORM} in more detail. Fig.~\ref{fig:ExSituSph} shows the mean \textit{ex situ} fraction of galaxy spheroids as a function of the spheroid mass fraction$-$the former is the mass fraction of the spheroid grown directly through mergers. We find that the \textit{ex situ} fraction generally grows with $M_\mathrm{*,sph}/M_\mathrm{*}$. However, the actual values of the \textit{ex situ} fraction are fairly low for almost all masses and $M_\mathrm{*,sph}/M_\mathrm{*}$ ratios. Even for massive galaxies ($M_* > 10^{11.25}$ M$_\odot$), mergers do not contribute a majority of the spheroid stellar mass, with the exception of pure spheroids ($M_\mathrm{*,sph}/M_\mathrm{*}\approx1$). Galaxies with $M_\mathrm{*,sph}/M_\mathrm{*}=0.8$ on average have less than $20\%$ of the spheroid stellar mass contributed by mergers. We find the following trends for intermediate and low-mass galaxies ($M_*<10^{10.5}$ M$_\odot$): 

i) A small increase in $M_\mathrm{*,sph}/M_\mathrm{*}$ from $M_\mathrm{*,sph}/M_\mathrm{*}=0$ is associated with a sharp rise in the \textit{ex situ} fraction. Even though the \textit{ex situ} fraction values are small, this means that mergers do play a significant role in creating spheroids for at most masses, even if they do not contribute greatly in terms of stellar mass.

ii) The \textit{ex situ} fraction falls with increasing stellar mass, at fixed values of $M_\mathrm{*,sph}/M_\mathrm{*}$. This trend is reversed only for massive galaxies ($M_* > 10^{11.25}$ M$_\odot$). This result can be associated with the importance of disc instabilities for intermediate-mass galaxies (e.g. \citealt{Lagos2008}, \citealt{Guo2011} and \citealt{Shankar2013}).

From our overall analysis it is clear that it is important to consider all channels of spheroid growth, alongside direct growth from mergers.

\subsection{Mergers vs. disc instabilities}

We will now consider the relative importance of all four channels of spheroid growth (separating the contributions of starbursts triggered by mergers and disc instabilities from the direct growth of spheroids due to both processes). We focus on the relative roles of these channels of growths in building up spheroids, regardless of the mass of the spheroids in question relative to their host galaxies.

\begin{figure}
\includegraphics[width=0.99\columnwidth, trim = 0 15 0 0]{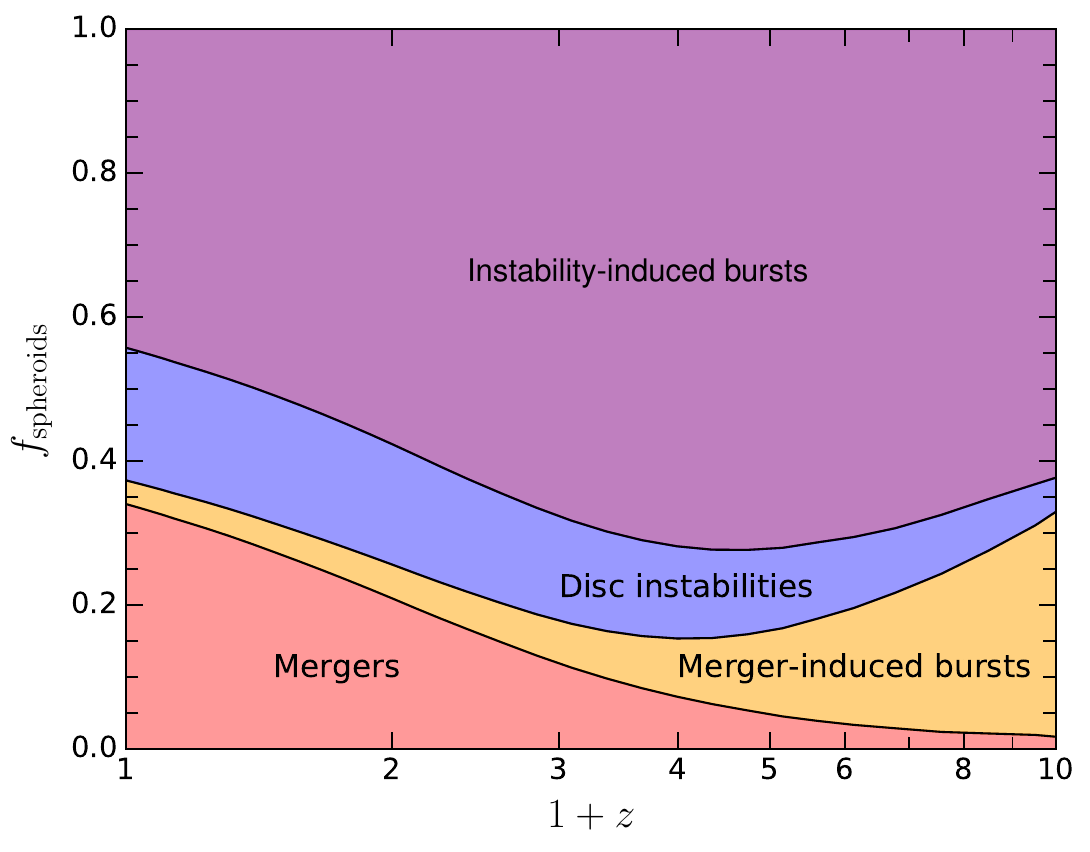}
\caption{Fractional contribution of individual channels of growth to the total stellar mass of all spheroids, as a function of redshift. Coloured regions show the fraction for each channel, as labeled.}
\label{fig:FracSphGlobal}
\end{figure}

We will first focus on the global contribution of each channel of growth to the total spheroid stellar mass density. In Fig.~\ref{fig:FracSphGlobal} we show the fractional contributions for each channel of mass growth, integrated over the spheroid mass function. In other words, this represents the fraction $M_{\mathrm{i,*,sph}}/M_\mathrm{*,sph}$ for a given channel of growth labeled as $i$, and $M_\mathrm{*,sph}$ is the total stellar mass for \textit{all} spheroids in the universe. Mergers provide a negligible contribution at very high redshifts, $z=10$, but their contribution increases at lower redshifts, with $35\%$ of all spheroid stellar mass provided through this channel. Merger-induced starbursts, on the other hand, show the opposite behaviour. Their contribution is only $5\%$ in the local Universe, but increases to $35\%$ at $z=10$. Disc instabilities provide $20\%$ of stellar mass in the local  universe and only $5\%$ at very high redshifts. Bursts triggered by disc instabilities dominate the total spheroid mass at all redshifts. At $z=0$ they provide $55\%$, peaking at $70\%$ at $z=4$, and falling down to $60\%$ at $z=10$.

\begin{figure*}
\includegraphics[width=0.99\textwidth, trim = 0 15 0 0]{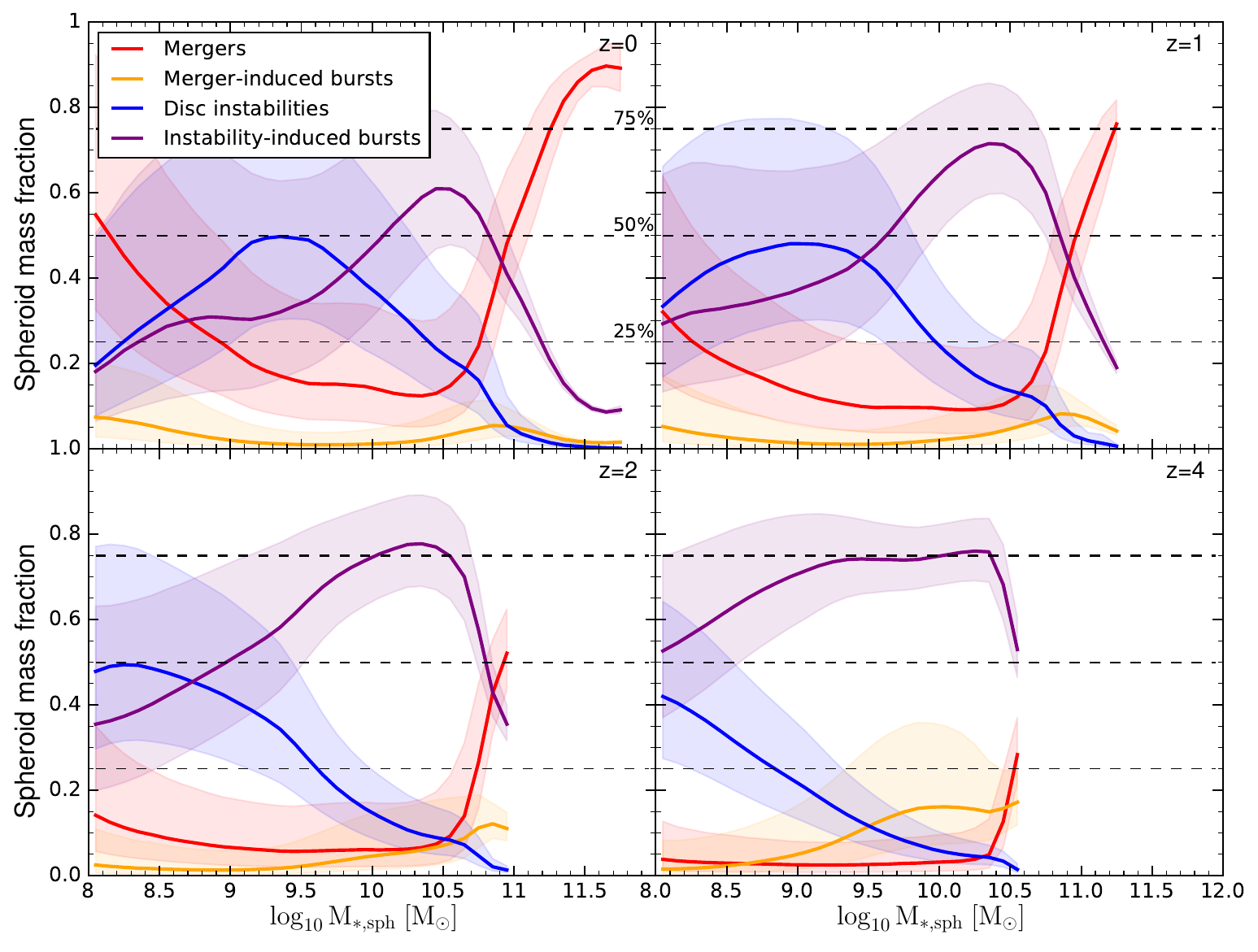}
\caption{Fractional contribution of individual channels of growth to the total stellar mass of spheroids (integrated over their entire history), as a function of spheroid mass. The contribution of each channel is given by lines coloured according to the legend. Shaded regions represent 16 to 84 percentile ranges. Each panel gives the dependencies at different redshifts, as labelled. }
\label{fig:FracSph}
\end{figure*}

It is interesting to consider how these contributions vary across different spheroid masses. Fig.~\ref{fig:FracSph} shows the fractional contribution of each channel of spheroid growth to the total spheroid stellar mass, as a function of spheroid stellar mass. At $z=0$ we find that mergers are almost solely responsible for the buildup of massive spheroids ($M_\mathrm{*,sph}>10^{11.3}$ M$_\odot$), contributing up to $90$ per cent of the mass. Surprisingly, with a contribution of $50$ per cent, they also play an important role in the buildup of very-low mass spheroids ($M_\mathrm{*,sph}<10^{8.5}$ M$_\odot$). This is due to disc instabilities being rarer for these spheroids than mergers (Fig.~\ref{fig:StarBurstInfo}). We find that merger-induced starbursts contribute an almost negligible mass for all spheroids ($0$-$7$ per cent).

Mergers contribute $10-20\%$ of the mass for intermediate mass spheroids  ($10^9<M_\mathrm{*,sph}<10^{11}$ M$_\odot$). We find that disc instabilities dominate for these spheroids. For $10^9<M_\mathrm{*,sph}<10^{10}$ M$_\odot$, the mass gained through disc destruction is larger than that created in the starbursts which arise from the disc instability. For $M_\mathrm{*,sph}>10^{10}$ M$_\odot$, however, the reverse is true. This is a result of the spheroid-to-total fraction growing significantly in this regime (\citealt{Lacey2016}), so that available disc masses are relatively small. However, these disc destructions are sufficient to trigger starbursts that begin to dominate the mass of the spheroid.

The scatter in the mean contribution from each channel is very large for all spheroids except the most massive ones. This illustrates the stochastic nature of the processes that drive the growth of spheroids. We find that the scatter is generally largest for low-mass spheroids ($M_\mathrm{*,sph}<10^{9}$ M$_\odot$), as a result of mergers becoming fairly important in this regime, as well as a small influence from merger-induced bursts. 

By considering the redshift evolution shown in the other panels of Fig.~\ref{fig:FracSph}, we find the following trends for the individual channels of spheroid growth:

i) Mergers become progressively less important with increasing redshift for most spheroids ($M_\mathrm{*,sph}<10^{10.5}$ M$_\odot$), but for higher mass spheroids 
we find a sharp rise in the fractional contribution of mergers, so much so that very massive spheroids ($M_\mathrm{*,sph}>10^{11}$ M$_\odot$) are built almost exclusively through mergers. This is true at lower redshifts, but is harder to confirm at higher redshifts due to a lack of such massive spheroids in our sample.

ii) Merger-induced starbursts contribute little to spheroid mass at all redshifts. At $z=4$ they are more important for the buildup of spheroids than mergers themselves, due to the strong decline in the direct contribution of mergers, but they are still much less important than disc instabilities.

iii) Disc instabilities generally have a larger influence at $z=0$ than at higher redshifts, except in low-mass spheroids. However, this is not a result of them being rare (in fact, they are more frequent at higher redshifts, see \ref{fig:StarBurstInfo}).

iv) Disc instability-induced starbursts become increasingly more important at higher redshifts, to the point of dominating by $z=4$,  for all spheroids in the mass range $M_\mathrm{*,sph}<10^{10.5}$ M$_\odot$. This is the result of disc instabilities being very frequent in the early universe. Since early discs are very gas-rich (e.g. \citealt{Daddi},~\citealt{Narayan}), this means that disc instabilities provide a constant supply of cold gas which fuels the central starbursts.

\subsection{Comparison with observations and other models}

\begin{figure*}
\includegraphics[width=0.99\textwidth, trim = 0 15 0 0]{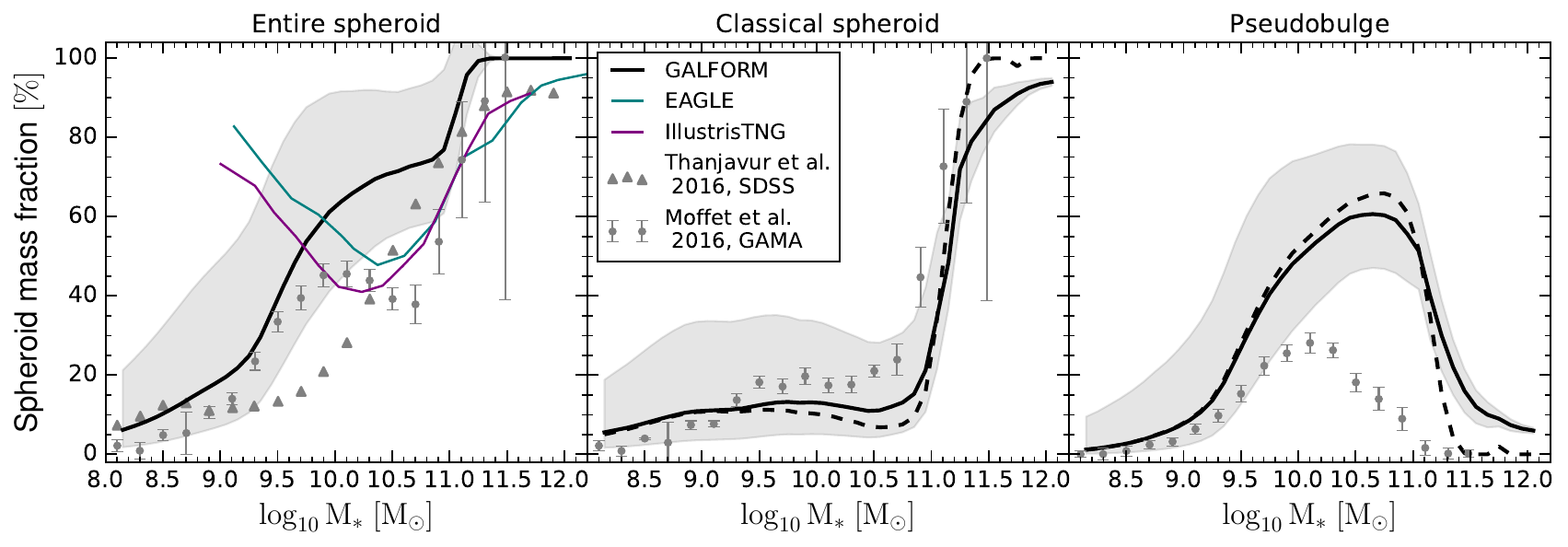}
\caption{The spheroid mass fraction (fraction of stellar mass in spheroids relative to the total stellar mass of a galaxy) as a function of stellar mass at $z=0$. Our results are shown with the thick black line (mean), with the shading indicating the 16 to 84 percentile range. We show the total spheroid mass fraction (\textit{left-hand panel}), the classical spheroid mass fraction (\textit{middle panel}, built by mergers) and the pseudobulge mass fraction (\textit{right-hand panel}, built by disc instabilities). These results are compared with observations from the SDSS and GAMA, as well as the EAGLE and Illustris-TNG simulations. The dashed black lines indicate classical spheroid and pseudobulge mass fractions obtained by classifying spheroids as entirely classical or entirely pseudobulge according to whether they were built mostly through mergers or disc instabilities, respectively$-$this definition matches the observational selection more closely.}
\label{fig:SphMassFracObs}
\end{figure*}

In the left-hand panel of Fig. \ref{fig:SphMassFracObs} we show the mean spheroid mass fraction in {\tt GALFORM} as a function of stellar mass at $z=0$, compared with observations from SDSS (\citealt{Thanjavur}) and GAMA (\citealt{Moffett}), as well as the EAGLE (\citealt{Clauwens}) and IllustrisTNG (\citealt{Tacchella2019}) hydrodynamical simulation. There is some disagreement between the SDSS and GAMA observations, but the {\tt GALFORM} spheroid mass fraction is higher than inferred from either of the surveys in the range $M_*=10^{9.5}-10^{11}$ $\mathrm{M}_\odot$. The hydrodynamical simulations straddle the observational data beyond $M_*=10^{10.5}$ $\mathrm{M}_\odot$, but they are too high for less massive galaxies$-$this disagreement is most likely due to the kinematic decomposition technique used for hydrodynamical simulations that tends to classify irregular galaxies as spheroidal (\citealt{Clauwens}).

In the middle and right-hand panels of Fig. \ref{fig:SphMassFracObs} we show the classical spheroid and pseudobulge mass fractions from {\tt GALFORM} compared with those from GAMA (\citealt{Moffett}). The latter were estimated by classifying spheroids as either diffuse or compact. From the middle panel we see that there is good agreement between {\tt GALFORM} and observations for the mass fraction in classical spheroids. On the other hand, the right-hand panel shows that the {\tt GALFORM} pseudobulge mass fraction is too high (by a factor of $\approx2$ at $M_*=10^{10}$ $\mathrm{M}_\odot$ and even more at higher masses), although the mass regime in which pseudobulges build up spheroids is roughly the same as observed. We have also attempted an alternative calculation of classical spheroid and pseudobulge fractions from {\tt GALFORM}, that more closely matches the observational method used by \cite{Moffett}. Namely, we classify spheroids as being entirely classical or entirely pseudobulges depending on whether they were built mostly by mergers or mostly by disc instabilities. These mass fractions are shown with dashed lines in the figures. They bring {\tt GALFORM} into better agreement with observations at very high masses ($M_*>10^{11}$ $\mathrm{M}_\odot$, but they do not bring down the peak pseudobulge mass fractions for intermediate-mass galaxies (on the contrary).

\cite{Parry2009} studied the contribution of mergers and disc instabilities to the buildup of spheroids, using older versions of {\tt GALFORM} (Durham model; \citealt{Bower2006}) and {\tt L-GALAXIES} (Munich model, \citealt{deLucia}). The contribution from disc instabilities was found to peak between $M_\mathrm{*}=10^{10}$ M$_\odot$ and $M_\mathrm{*}=10^{11}$ M$_\odot$ for both models. 
However, the peak value for {\tt L-GALAXIES} was $40\%$ instead of the $70\%$ predicted by {\tt GALFORM}. In the current version of {\tt GALFORM} we find a peak of $60\%$ (due to disc instabilities directly as well as their starbursts), at roughly $M_\mathrm{*}=10^{10.7}$ M$_\odot$. This is close to the value found by \cite{Parry2009}, and in the same mass range.

In addition to mass fractions, observations have often focused on the pseudobulge fraction$-$the fraction of all galaxies that host a pseudobulge (as opposed to a classical spheroid, or no spheroid at all). For this purpose we classify a spheroid as a pseudobulge if it formed at least half of its stellar mass through disc instabilities (including starbursts triggered by disc instabilities). Spheroids that do not match this definition are classified as classical. Galaxies whose spheroids have a very low mass compared to the disc ($M_\mathrm{*,sph}/M_\mathrm{*}<0.01$) are labeled as bulgeless. We then calculate the pseudobulge fraction as the fraction of galaxies (in a given stellar mass or redshift bin) that host a pseudobulge. The definitions we have used here are in line with previous theoretical studies of pseudobulge fractions (\citealt{Shankar2013},~\citealt{Izquierdo-Villalba2019}).

\begin{figure*}
\includegraphics[width=0.99\textwidth, trim = 0 15 0 0]{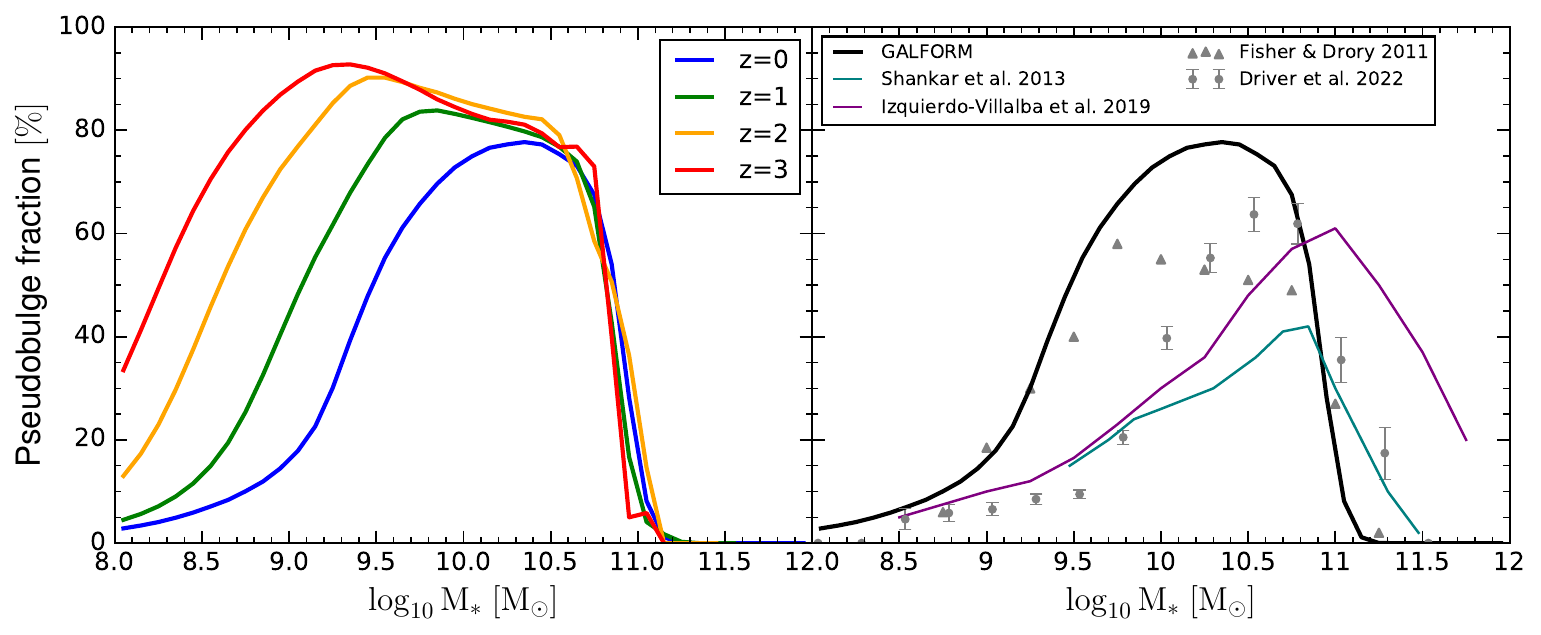}
\caption{Pseudobulge fraction as a function of stellar mass, at several redshifts (\textit{left-hand panel}), and compared to observations and other semi-analytical models at $z=0$ (\textit{right-hand panel}). Pseudobulges are defined as any spheroids within a galaxy with $M_\mathrm{*,sph}/M_\mathrm{*}>0.01$ that were built more through disc instabilities than mergers (including starbursts caused by both mechanisms). The pseudobulge fraction is then calculated as the fraction of all galaxies at a given stellar mass that host such a spheroid.}
\label{fig:FracPseudo}
\end{figure*}

In the left-hand panel of Fig. \ref{fig:FracPseudo} we show the pseudobulge fraction in {\tt GALFORM} at several redshifts. At $z=0$ we find that the pseudobulge fraction peaks between $M_*=10^{9.5}$ M$_\odot$ and $M_*=10^{11}$ M$_\odot$, with a peak value $\approx80\%$. Beyond $M_*=10^{11}$ M$_\odot$ the pseudobulge fraction drops sharply due to the increasing impact of mergers. The drop at lower masses is also caused by mergers, but this drop is more protracted. The drop at higher masses is possibly sharper due to AGN feedback that sharply turns off star formation, and the growth of galaxy discs, in such galaxies. We find that at lower masses ($M_*<10^{10}$ M$_\odot$), the pseudobulge fraction increases with redshift, showing that mergers are less important for these galaxies at earlier times. The sharp drop at $M_*=10^{11}$ M$_\odot$ is independent of redshift. The peak in the pseudobulge fraction between $M_*=10^{9.5}$ M$_\odot$ and $M_*=10^{10.5}$ M$_\odot$ rises slightly with redshift up to $z=2$, but then begins to drop due to the increasing influence of merger-induced starbursts at higher redshifts (see Fig. \ref{fig:FracSphGlobal} or \ref{fig:FracSph}).

In the right-hand panel of Fig. \ref{fig:FracPseudo} we compare against a few theoretical models and observational estimates at $z=0$. \cite{Fisher2011} used a large sample of galaxies observed in the local Universe to calculate a pseudobulge fraction. They classified their spheroids morphologically; any spheroid with $M_\mathrm{*,sph}/M_\mathrm{*}>0.01$ and a Sérsic index of $n<2$ was classified as a pseudobulge. They find significant pseudobulge fractions in the same population of galaxies as we do (i.e. the peak in our pseudobulge fraction has a similar shape), but with lower peak values ($60\%$ versus our $80\%$). We also compare with data from GAMA (\citealt{Driver2022}) that is based on morphological classifications$-$we assume that their 'compact bulge with disc' galaxy type matches our pseudobulge selection. Our pseudobulge fraction is higher than based on these data, except for a narrow range of stellar masses: $M_*=10^{10.5}-10^{11}$ $\mathrm{M}_\odot$. However, the pseudobulge fraction from \cite{Driver2022} is also higher than the \cite{Fisher2011} one in the same range, and lower for $M_*<10^{10}$ $\mathrm{M}_\odot$. ~\cite{Vaghamere2013} studied a small sample of intermediate mass galaxies, finding 77 per cent to host pseudobulges, in better agreement with our predictions. \cite{Erwin2015} also used an intermediate mass sample ($M_\mathrm{*}\approx10^{10}$ M$_\odot$) and decomposed spheroids into their classical and pseudobulge components. They found that the classical spheroid component contributes only $6$ per cent to galaxy stellar masses, on average, while the pseudobulge component contributes $11$-$59$ per cent. 

\cite{Shankar2013} used an updated version of the Munich model, and found a fraction of pseudobulges that peaks at $M_\mathrm{*}=10^{10.8}$ M$_\odot$, with a decrease towards both lower and higher masses. These results are in qualitative agreement with ours, but as can be seen from Fig. \ref{fig:FracPseudo}, the position and shape of the peak, as well as its values, are different from {\tt GALFORM}. Their peak pseudobulge fractions are lower, are reached at a higher stellar mass, and in a narrower range of stellar masses. \cite{Izquierdo-Villalba2019} studied the pseudobulge fraction using the most recent version of {\tt L-GALAXIES}. They found an overall similar result as \cite{Shankar2013}, but the peak in the pseudobulge fraction is higher (yet still lower than predicted by {\tt GALFORM}).

Overall we find that {\tt GALFORM} has a relatively high pseudobulge fraction, regardless of whether we define it as a mass fraction (Fig. \ref{fig:SphMassFracObs}) or as a morphological type fraction (Fig. \ref{fig:FracPseudo}). This is true when we compare our predictions with both observations and other models. These comparisons indicate that the modeling of disc instabilities in {\tt GALFORM} may be somewhat unrealistic. In particular, {\tt GALFORM} assumes that once a disc instability is triggered, the entire disc is destroyed and transformed/transferred to the spheroid. This assumption is somewhat extreme and represents an 'upper limit' of the effects of disc instabilities.

There are other modifications to the treatment of disc instabilities in {\tt GALFORM} that may lead to better agreement with observations. In particular, the transferral of the disc component to the bulge may need to be modeled as occurring over some time-scale instead of being instantaneous (corresponding to secular evolution). In addition, the disc instability mode that features clump migration in gas-rich discs, rather than bar formation, may need to be included. The stability criterion, which determines when disc instabilities are triggered, may also need to be modified. In future work we intend to study which of these modifications need to be made, and what are the most appropriate parameters that would yield better agreement between {\tt GALFORM} and observations.

\subsection{Star formation in bursts: mergers vs. disc instabilities}
\label{sec:StarFormationBursts}

From the results outlined in previous subsections, it is clear that star formation in bursts is an important mechanism of spheroid growth. This is especially true for disc instability-induced bursts. Here we will look at the star formation rates in bursts directly.

Eqn.~\ref{eq:BurstSFR} shows that the SFR in these bursts depends on the cold gas mass transferred to the spheroid, as well as the dynamical timescale of the bulge after the cold gas has been transferred to it. Furthermore, in order to calculate the average SFR across all spheroids, knowledge of the frequency of mergers and disc instabilities is required. We discuss these in detail in Appendix~\ref{app1}.

Fig.~\ref{fig:SFRBursts} shows the predicted specific SFR (sSFR) as a function of spheroid mass at several redshifts. The left-hand panel shows the sSFR in ongoing bursts only, for mergers and disc instabilities. Bursts in both modes result in  sSFRs that fall with increasing spheroid mass. This is related to the decrease of the gas fractions at higher spheroid masses, as well as the increase in the burst timescale. We find that merger-induced bursts have a much lower SFR, at least a factor of 10 at $z=0$, depending on spheroid mass. This is consistent with our results on mass fractions from individual channels of spheroid growth (see previous subsections). 

The huge difference between merger and disc instability-induced starbursts is a result of two effects: i) post-merger bursts host lower amounts of cold gas, and ii) spheroids created by mergers are significantly larger, which translates into a much longer burst timescale. At higher redshifts the difference is smaller, as a result of cold gas masses in mergers becoming comparable to those in disc instabilities. In addition, the burst timescales in disc instabilities rise at higher redshifts, while those in mergers fall. By $z=4$ this results in the merger-induced sSFRs being only a factor of two (at most) lower than those induced by disc instabilities.

\begin{figure*}
\includegraphics[width=0.99\textwidth, trim = 0 15 0 0]{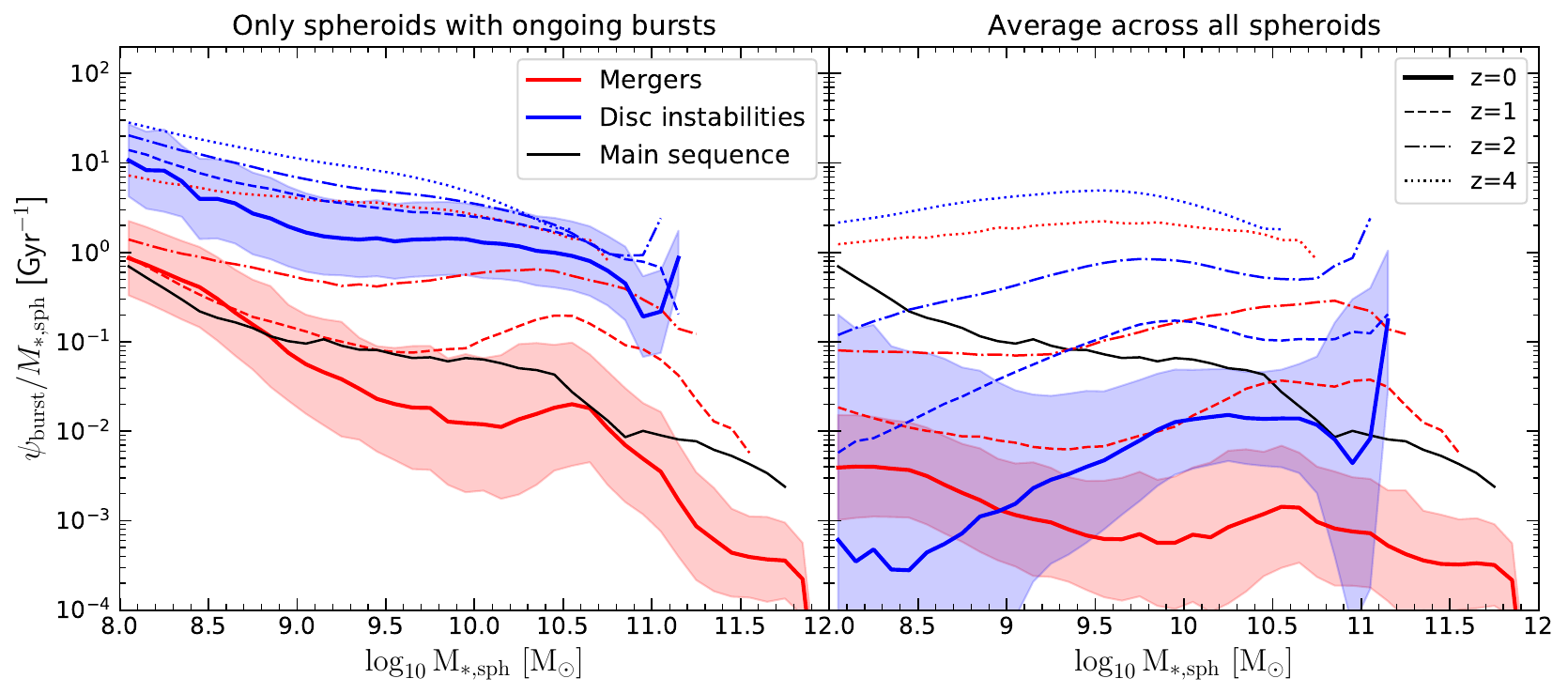}
\caption{Specific star formation rates in starbursts as a function of spheroid stellar mass. The sSFR caused by mergers and disc instabilities are shown by red and blue lines, respectively, while the main sequence (non-burst and non-passive) sSFR is shown by the black line for $z=0$. Different line types denote different redshifts, as per the legend in the right-hand panel. The shading indicates 1$\sigma$ dispersion around the mean. \textit{Left:} sSFR in spheroids that currently host an ongoing starburst. \textit{Right:} Average sSFR across all spheroids, calculated by including spheroids with $\psi_\mathrm{burst}=0$.}
\label{fig:SFRBursts}
\end{figure*}

While these differences are interesting to consider on their own, the relative role of these two channels of star formation cannot be gauged without knowing how often they occur. With this in mind, we calculate the \textit{average} sSFR by including the frequency of mergers and disc instabilities at a given spheroid mass, as well as by taking into account the spheroid mass function. In the right-hand panel of Fig.~\ref{fig:SFRBursts}, we show this average sSFR. Disc instabilities still dominate across most spheroid masses. Mergers dominate for low-mass spheroids ($M_\mathrm{*,sph}<10^{9}$ M$_\odot$ at $z=0$, $M_\mathrm{*,sph}<10^{8}$ M$_\odot$ at $z=2$. This is a result of disc instabilities being relatively rare at these masses (see Fig.~\ref{fig:StarBurstInfo}.

In Fig. ~\ref{fig:SFRBursts} we also show the main sequence specific star formation rate for $z=0$, calculated as the average sSFR of all galaxies that are active (sSFR$>0.01$ Gyr$^{-1}$) and that are currently not hosting a starburst. The starburst sSFR becomes comparable to the main sequence one at $M_*=10^{11}$ M$_\odot$, due to the increasing effect of disc instabilities. Here we will not compare our SFRs with observed ones, as this is beyond the scope of this paper. See \cite{Cowley} for a detailed discussion of main sequence and starburst SFRs in {\tt GALFORM}.

\section{Summary and conclusions}
\label{sec:Conclusion}

We used the {\tt GALFORM} semi-analytical model, set in the Planck Millennium N-body simulation, to study the growth of the stellar content of galaxies and their spheroids. We investigated the importance of mergers in the overall growth through the merger accretion rate and the \textit{ex situ} mass fraction. We explored the dependence of these quantities  on stellar mass and redshift. Our findings can be summarised as follows.

\begin{itemize}
\item The accretion of stellar mass through  mergers dominates over \textit{in situ} star formation for galaxies more massive than $M_*>10^{10.7}$ M$_\odot$ at $z=0$. This transition mass increases with redshift, and reaches $M_*=10^{11}$ M$_\odot$ by $z=2$. Our predictions for massive galaxies are consistent with observational constraints. At very high masses ($M_*=10^{11.5}$ M$_\odot$ at $z=0$) the mass growth rate due to mergers begins to decline as a result of the decline in galaxy numbers seen in the galaxy stellar mass function.
\item The \textit{ex situ} fraction of galaxies increases sharply for massive galaxies ($M_*>10^{11}$ M$_\odot$), reaching $90\%$ by $M_*=10^{11.5}$ M$_\odot$ at $z=0$, in agreement with observations. This transition, and the values of the \textit{ex situ} fraction for such galaxies, remain constant with redshift. This lack of evolution with redshift is in disagreement with other published models. {\tt GALFORM} also predicts a higher \textit{ex situ} fraction for low-mass galaxies ($M_*<10^{10.7}$ M$_\odot$) than other models, on the order of $7$-$12$ per cent. These values decrease with redshift, again in disagreement with other models.
\item Major mergers ($\mu_*\in[0.25,1]$) contribute roughly $50$ per cent to the total \textit{ex situ} stellar mass of galaxies, with minor mergers ($\mu_*\in[0.1,0.25]$) and accretion ($\mu_*<0.1$) each contributing $25$ per cent. These fractions evolve weakly with redshift. Massive galaxies ($M_*=10^{11}$ M$_\odot$) display the largest mass growth contribution from major mergers, at $80$ per cent, but very massive galaxies ($M_*>10^{11.5}$ M$_\odot$) have a decreasing fraction grown from major mergers. This is the result of a lack of similar-mass companions for these largest cluster galaxies, which are constrained to grow further only through relatively minor mergers.
\item The global \textit{ex situ} fraction, defined as the contribution of \textit{ex situ} mass to the total stellar mass of the universe, evolves strongly with redshift. $40$ per cent of the stars in the local universe are predicted to reside in galaxies in which they did not form. This fraction falls to $3$ per cent by $z=4$. The global fraction of $40$ per cent at $z=0$ is made up of  $29$ per cent through major mergers, $6.5$ per cent from minor mergers and $4.5$ per cent by accretion.
\end{itemize}

\noindent
In the second part of our analysis we focused on the growth of spheroids. We tracked the history of mergers and disc instabilities for individual spheroids in order to calculate the stellar masses associated with these events. We calculated spheroid mass fractions and compared the relative contributions to the growth of spheroids from both processes, as well as the starbursts that they trigger. We calculated star formation rates in bursts caused by both mergers and disc instabilities. We conclude the following:

\begin{itemize}
\item The spheroid mass fraction grows monotonically with stellar mass at $z=0$, due to a complex interplay between mergers and disc instabilities. Low-mass galaxies ($M_*<10^9$ $\mathrm{M}_\odot$) have a low, but non-negligible spheroid mass fraction from mergers. Between $M_*=10^9$ $\mathrm{M}_\odot$ and $M_*=10^{11}$ $\mathrm{M}_\odot$, disc instabilities dominate, with a contribution of up to $60\%$. Mergers again dominate for $M_*>10^{11}$ $\mathrm{M}_\odot$, where the spheroids they create also dominate the total stellar mass of the galaxies. Merger-induced starburst are negligible at all but the highest redshifts.
\item The dominant spheroid mass fraction from disc instabilities in the intermediate mass range ($10^{8.5}<M_\mathrm{*,sph}<10^{11}$ M$_\odot$) is in agreement with the hypothesis that they are the main cause of pseudobulge growth. In this regime they contribute both directly and through starbursts in approximately equal proportions. The direct channel dominates for lower-mass spheroids (peaking at $M_\mathrm{*,sph}=10^{9.5}$ M$_\odot$), while starbursts dominate for higher-mass spheroids (peaking at $M_\mathrm{*,sph}=10^{10.5}$ M$_\odot$). At higher redshifts, we predict an increasing fraction due to disc instability-induced starbursts. At $z=4$, they dominate for all galaxies but the most massive ones. 
\item Integrating these contributions over the spheroid mass function, we find the global fractions of spheroid mass grown through each of these channels. Mergers contribute $35$ per cent of all spheroid mass at $z=0$, reducing monotonically to almost zero at $z=10$. The starbursts they cause, on the other hand, provide only $5$ per cent of the mass at $z=0$, but $35$ per cent at $z=10$. Disc instabilities grow from $5$ per cent at $z=10$ to $20$ per cent in the local universe. Disc instability-induced starbursts dominate overall, contributing more than $50$ per cent of the mass at all redshifts. Their contribution peaks at $70$ per cent at $z=4$, with merger-induced starbursts becoming increasingly important in the early universe.
\item The modeling of disc instabilities is likely too extreme in {\tt GALFORM}, leading to pseudobulge fractions (including mass fractions and morphological type fractions) that exceed observations in the peak value. 
\item The specific star formation rate (sSFR) in ongoing starbursts at $z=0$ is much larger for disc instabilities than it is for mergers (by at least a factor of 10, depending on mass). Both merger and disc instability-induced starbursts have higher sSFRs at higher redshifts, with merger-induced bursts growing more sharply. When averaged over all spheroids, we find that the difference is less stark, with merger-induced starbursts dominating for low-mass spheroids ($M_*<10^{9}$ M$_\odot$).
\end{itemize}

We provide quantitative predictions for the importance of mergers, disc instabilities and their starbursts for both the local and distant Universe. Upcoming high-redshift observatories, such as the JWST, will be able to test these predictions in more detail. In particular, the frequency of mergers and the SFR in ongoing bursts at high redshifts will be able to distinguish which processes are most important for the growth of spheroids, as well as that of their host galaxies. The cosmological evolution of classical and pseudobulge fractions will also help further constrain the treatment of mergers and disc instabilities in models of galaxy formation.




\section*{Acknowledgements}
F. H. would like to acknowledge support from the Science Technology Facilities Council through a CDT studentship (ST/P006744/1). This work was also supported by STFC grant ST/T000244/1. This work used the DiRAC@Durham facility managed by the Institute for Computational Cosmology on behalf of the STFC DiRAC HPC Facility (www.dirac.ac.uk). The equipment was funded by BEIS capital funding via STFC capital grants ST/K00042X/1, ST/P002293/1, ST/R002371/1 and ST/S002502/1, Durham University and STFC operations grant ST/R000832/1. DiRAC is part of the National e-Infrastructure.




\section*{Data availability}

The data underlying this article will be provided upon request to the corresponding author.

\bibliographystyle{mnras}
\bibliography{mnras_template} 




\appendix

\section{Starbursts in spheroids - mergers vs. disc instabilities}
\label{app1}

Here we discuss the behaviour of various factors that contribute to the SFR in bursts shown in Fig.~\ref{fig:SFRBursts}. As shown by Equation~\ref{eq:BurstSFR}, these are the cold gas mass in ongoing bursts and a timescale which is related to the dynamical timescale of the spheroid. We consider both of these properties for mergers and disc instabilities. In order to calculate the average SFR across all spheroids (including those that are not under going starbursts), we also need the relative frequency of mergers and disc instabilities. We choose to quantify this through merger and disc instability rate densities.

\begin{figure}
\includegraphics[width=0.99\columnwidth, trim = 0 15 0 0]{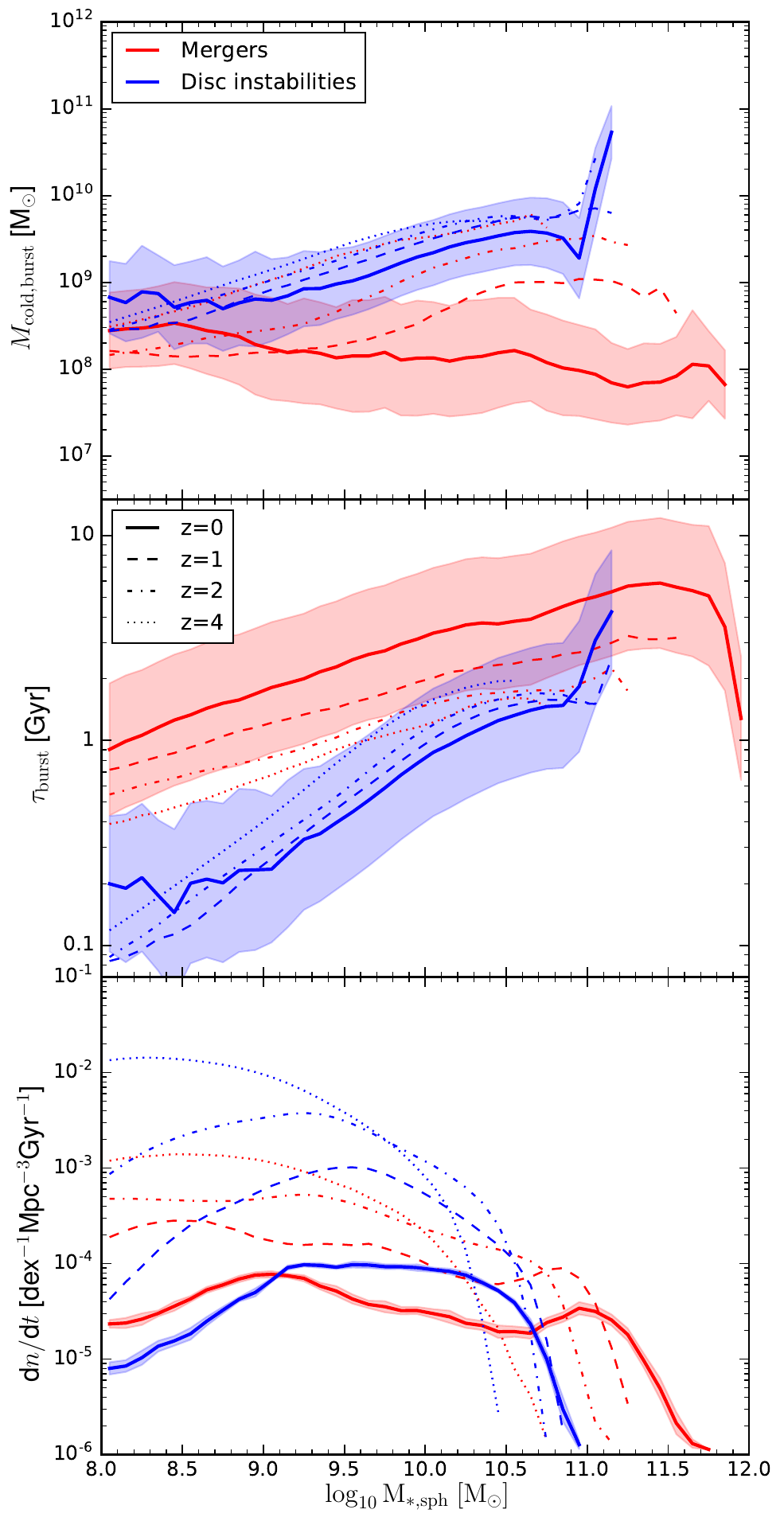}
\caption{Ingredients for the calculation of the SFR in bursts, as functions of spheroid stellar mass. Red and blue lines show quantities relating to mergers and disc instabilities, respectively. Different line types show relations at different redshifts, as indicated by the key. Shaded regions represent the 16 to 84 percentile ranges. \textit{Top:} Cold gas mass in ongoing bursts. \textit{Middle:} timescale used in the burst SFR calculation. \textit{Bottom:} Merger and disc instability rate densities.} 
\label{fig:StarBurstInfo}
\end{figure}

The top panel of Fig.~\ref{fig:StarBurstInfo} shows the cold gas mass in currently ongoing star formation bursts ($M_\mathrm{cold}$) as a function of spheroid stellar mass ($M_\mathrm{*,sph}$). This is shown separately for starbursts caused by mergers and disc instabilities. We find that $M_\mathrm{cold}$ is larger for disc instability starbursts, regardless of $M_\mathrm{*,sph}$ or redshift. For disc instabilities, $M_\mathrm{cold}$ rises with $M_\mathrm{*,sph}$ at all redshifts, and at a similar pace. There is also a small rise with redshift in the overall relation for disc instability-induced starbursts. Merger-induced starbursts at $z=0$ contain a decreasing amount of $M_\mathrm{cold}$ as a function of spheroid mass. By $z=1$ this is no longer true, with $M_\mathrm{cold}$ growing at a similar pace as for disc instabilities. We find that an increase in redshift leads to a more drastic rise in $M_\mathrm{cold}$ for mergers, so that by $z=4$  starbursts induced by mergers and disc instabilities host similar amounts of cold gas. 

The middle panel of Fig.~\ref{fig:StarBurstInfo} shows the variation of the starburst timescale with spheroid mass at different redshifts. Both for mergers and disc instabilities, this timescale grows as a function of $M_\mathrm{*,sph}$, except for the most massive spheroids formed through mergers. 
Merger-induced starbursts have significantly longer timescales than disc instability-induced ones at $z=0$. 
By $z=4$, this difference is much smaller, owing to merger-induced timescales falling with redshift, and disc instability-induced ones rising with redshift. 
These difference are not intuitive, but can likely be tracked down to the timescale that governs starbursts (see Eqn.\ref{eq:BurstTimeScale}), which is primarily determined by the size of the spheroid. Spheroid sizes are in turn governed by several mechanisms, among which are  mergers and disc instabilities, but also adiabatic relaxation in the gravitational potential of the disc, bulge and dark matter halo. 

The relative frequency of mergers and disc instabilities is important when considering the impact of either process on the buildup of spheroids. In practice, the calculation of the \textit{average} SFR in bursts entails summing the burst SFR over all spheroids, including those with $\psi_\mathrm{burst}=0$. For the purposes of displaying this information, we instead calculate the rate of either mergers or disc instabilities. This then yields the usual merger rate density (\citealt{Lotz2011}, \citealt{Xu2012}, \citealt{Stott2013}), and a disc instability rate density. The bottom panel of Fig.~\ref{fig:StarBurstInfo} shows the merger and disc instability rate densities as functions of $M_\mathrm{*,sph}$ and $z$. Mergers dominate for massive spheroids at all redshifts ($M_\mathrm{*,sph}>10^{10.7}$ M$_\odot$ at $z=0$, $M_\mathrm{*,sph}>10^{10.4}$ M$_\odot$ at $z=4$). They also dominate for low-mass spheroids, but only at low redshifts. The delimiting mass moves from $M_\mathrm{*,sph}=10^{9}$ M$_\odot$ at $z=0$ to $M_\mathrm{*,sph}=10^{8}$ M$_\odot$ at $z=4$. Disc instabilities are more frequent than mergers by a factor of several in the intermediate mass regime ($10^{9}<M_\mathrm{*,sph}<10^{10.5}$ M$_\odot$). In combination with the trends of the SFR in ongoing bursts (left-hand panel of Fig.~\ref{fig:SFRBursts}), this means that we expect disc instability bursts to dominate the mass budget of intermediate mass spheroids at all redshifts. This behaviour is confirmed in the averaged SFR, shown in the right-hand panel of Fig.~\ref{fig:SFRBursts}.



\bsp	
\label{lastpage}
\end{document}